\documentclass[longauth]{aa}

\usepackage{euclid}
\usepackage{graphicx}
\usepackage{natbib}
\usepackage{scalerel}
\usepackage{multirow}   
\usepackage{mathrsfs}
\usepackage[table]{xcolor}
\usepackage{fontawesome}

\newcommand{\og}{{\tt OpenGADGET3}}
\newcommand{\pinocchio}{{\tt PINOCCHIO}}
\newcommand{\subfind}{{\tt SUBFIND}}
\newcommand{\piccolo}{{\tt PICCOLO}}
\newcommand{\rockstar}{{\tt ROCKSTAR}}
\newcommand{\consistent}{{\tt CONSISTENT}}
\newcommand{\monofonic}{{\tt monofonIC}}
\newcommand{\pylians}{{\tt PYLIANS}}
\newcommand{\pymc}{{\tt PyMC}}
\newcommand{\msun}{M_{\odot}\,h^{-1}}
\newcommand{\mpc}{h^{-1}\,\textup{Mpc}}
\newcommand{\mpcinv}{h\,\textup{Mpc}^{-1}}
\newcommand{\de}{\text{d}}

\usepackage{txfonts}
\usepackage[pdfencoding=auto,psdextra]{hyperref}
\hypersetup{
    colorlinks=true,
    linkcolor=blue,
    filecolor=magenta,      
    urlcolor=blue,
    citecolor=blue
}
\makeatletter
\renewcommand*\aa@pageof{, page \thepage{} of \pageref*{LastPage}}
\makeatother

\usepackage[utf8]{inputenc}

\usepackage[switch, modulo]{lineno}

\begin{document}
%
%
\title{\Euclid preparation}
\subtitle{L. Calibration of the halo linear bias in $\Lambda(\nu)$CDM cosmologies}    
\newcommand{\orcid}[1]{} 
\author{Euclid Collaboration: T.~Castro\orcid{0000-0002-6292-3228}\thanks{\email{tiago.batalha@inaf.it}}\inst{\ref{aff1},\ref{aff2},\ref{aff3},\ref{aff4}}
\and A.~Fumagalli\orcid{0009-0004-0300-2535}\inst{\ref{aff5},\ref{aff3}}
\and R.~E.~Angulo\orcid{0000-0003-2953-3970}\inst{\ref{aff6},\ref{aff7}}
\and S.~Bocquet\orcid{0000-0002-4900-805X}\inst{\ref{aff8}}
\and S.~Borgani\orcid{0000-0001-6151-6439}\inst{\ref{aff9},\ref{aff3},\ref{aff1},\ref{aff2}}
\and M.~Costanzi\orcid{0000-0001-8158-1449}\inst{\ref{aff9},\ref{aff1},\ref{aff3}}
\and J.~Dakin\orcid{0000-0002-2915-0315}\inst{\ref{aff10}}
\and K.~Dolag\inst{\ref{aff8}}
\and P.~Monaco\orcid{0000-0003-2083-7564}\inst{\ref{aff9},\ref{aff1},\ref{aff2},\ref{aff3}}
\and A.~Saro\orcid{0000-0002-9288-862X}\inst{\ref{aff9},\ref{aff3},\ref{aff1},\ref{aff2},\ref{aff4}}
\and E.~Sefusatti\orcid{0000-0003-0473-1567}\inst{\ref{aff1},\ref{aff3},\ref{aff2}}
\and N.~Aghanim\orcid{0000-0002-6688-8992}\inst{\ref{aff11}}
\and L.~Amendola\orcid{0000-0002-0835-233X}\inst{\ref{aff12}}
\and S.~Andreon\orcid{0000-0002-2041-8784}\inst{\ref{aff13}}
\and C.~Baccigalupi\orcid{0000-0002-8211-1630}\inst{\ref{aff3},\ref{aff1},\ref{aff2},\ref{aff14}}
\and M.~Baldi\orcid{0000-0003-4145-1943}\inst{\ref{aff15},\ref{aff16},\ref{aff17}}
\and C.~Bodendorf\inst{\ref{aff18}}
\and D.~Bonino\orcid{0000-0002-3336-9977}\inst{\ref{aff19}}
\and E.~Branchini\orcid{0000-0002-0808-6908}\inst{\ref{aff20},\ref{aff21},\ref{aff13}}
\and M.~Brescia\orcid{0000-0001-9506-5680}\inst{\ref{aff22},\ref{aff23},\ref{aff24}}
\and A.~Caillat\inst{\ref{aff25}}
\and S.~Camera\orcid{0000-0003-3399-3574}\inst{\ref{aff26},\ref{aff27},\ref{aff19}}
\and V.~Capobianco\orcid{0000-0002-3309-7692}\inst{\ref{aff19}}
\and C.~Carbone\orcid{0000-0003-0125-3563}\inst{\ref{aff28}}
\and J.~Carretero\orcid{0000-0002-3130-0204}\inst{\ref{aff29},\ref{aff30}}
\and S.~Casas\orcid{0000-0002-4751-5138}\inst{\ref{aff31}}
\and M.~Castellano\orcid{0000-0001-9875-8263}\inst{\ref{aff32}}
\and G.~Castignani\orcid{0000-0001-6831-0687}\inst{\ref{aff16}}
\and S.~Cavuoti\orcid{0000-0002-3787-4196}\inst{\ref{aff23},\ref{aff24}}
\and A.~Cimatti\inst{\ref{aff33}}
\and C.~Colodro-Conde\inst{\ref{aff34}}
\and G.~Congedo\orcid{0000-0003-2508-0046}\inst{\ref{aff35}}
\and C.~J.~Conselice\orcid{0000-0003-1949-7638}\inst{\ref{aff36}}
\and L.~Conversi\orcid{0000-0002-6710-8476}\inst{\ref{aff37},\ref{aff38}}
\and Y.~Copin\orcid{0000-0002-5317-7518}\inst{\ref{aff39}}
\and A.~Costille\inst{\ref{aff25}}
\and F.~Courbin\orcid{0000-0003-0758-6510}\inst{\ref{aff40}}
\and H.~M.~Courtois\orcid{0000-0003-0509-1776}\inst{\ref{aff41}}
\and A.~Da~Silva\orcid{0000-0002-6385-1609}\inst{\ref{aff42},\ref{aff43}}
\and H.~Degaudenzi\orcid{0000-0002-5887-6799}\inst{\ref{aff44}}
\and G.~De~Lucia\orcid{0000-0002-6220-9104}\inst{\ref{aff1}}
\and A.~M.~Di~Giorgio\orcid{0000-0002-4767-2360}\inst{\ref{aff45}}
\and M.~Douspis\orcid{0000-0003-4203-3954}\inst{\ref{aff11}}
\and X.~Dupac\inst{\ref{aff38}}
\and S.~Dusini\orcid{0000-0002-1128-0664}\inst{\ref{aff46}}
\and M.~Farina\orcid{0000-0002-3089-7846}\inst{\ref{aff45}}
\and S.~Farrens\orcid{0000-0002-9594-9387}\inst{\ref{aff47}}
\and S.~Ferriol\inst{\ref{aff39}}
\and P.~Fosalba\orcid{0000-0002-1510-5214}\inst{\ref{aff48},\ref{aff49}}
\and M.~Frailis\orcid{0000-0002-7400-2135}\inst{\ref{aff1}}
\and E.~Franceschi\orcid{0000-0002-0585-6591}\inst{\ref{aff16}}
\and M.~Fumana\orcid{0000-0001-6787-5950}\inst{\ref{aff28}}
\and S.~Galeotta\orcid{0000-0002-3748-5115}\inst{\ref{aff1}}
\and B.~Gillis\orcid{0000-0002-4478-1270}\inst{\ref{aff35}}
\and C.~Giocoli\orcid{0000-0002-9590-7961}\inst{\ref{aff16},\ref{aff50}}
\and P.~G\'omez-Alvarez\orcid{0000-0002-8594-5358}\inst{\ref{aff51},\ref{aff38}}
\and A.~Grazian\orcid{0000-0002-5688-0663}\inst{\ref{aff52}}
\and F.~Grupp\inst{\ref{aff18},\ref{aff8}}
\and L.~Guzzo\orcid{0000-0001-8264-5192}\inst{\ref{aff53},\ref{aff13}}
\and S.~V.~H.~Haugan\orcid{0000-0001-9648-7260}\inst{\ref{aff54}}
\and W.~Holmes\inst{\ref{aff55}}
\and F.~Hormuth\inst{\ref{aff56}}
\and A.~Hornstrup\orcid{0000-0002-3363-0936}\inst{\ref{aff57},\ref{aff58}}
\and S.~Ili\'c\orcid{0000-0003-4285-9086}\inst{\ref{aff59},\ref{aff60}}
\and K.~Jahnke\orcid{0000-0003-3804-2137}\inst{\ref{aff61}}
\and M.~Jhabvala\inst{\ref{aff62}}
\and B.~Joachimi\orcid{0000-0001-7494-1303}\inst{\ref{aff63}}
\and E.~Keih\"anen\orcid{0000-0003-1804-7715}\inst{\ref{aff64}}
\and S.~Kermiche\orcid{0000-0002-0302-5735}\inst{\ref{aff65}}
\and A.~Kiessling\orcid{0000-0002-2590-1273}\inst{\ref{aff55}}
\and M.~Kilbinger\orcid{0000-0001-9513-7138}\inst{\ref{aff47}}
\and B.~Kubik\orcid{0009-0006-5823-4880}\inst{\ref{aff39}}
\and M.~Kunz\orcid{0000-0002-3052-7394}\inst{\ref{aff66}}
\and H.~Kurki-Suonio\orcid{0000-0002-4618-3063}\inst{\ref{aff67},\ref{aff68}}
\and P.~B.~Lilje\orcid{0000-0003-4324-7794}\inst{\ref{aff54}}
\and V.~Lindholm\orcid{0000-0003-2317-5471}\inst{\ref{aff67},\ref{aff68}}
\and I.~Lloro\inst{\ref{aff69}}
\and E.~Maiorano\orcid{0000-0003-2593-4355}\inst{\ref{aff16}}
\and O.~Mansutti\orcid{0000-0001-5758-4658}\inst{\ref{aff1}}
\and O.~Marggraf\orcid{0000-0001-7242-3852}\inst{\ref{aff70}}
\and K.~Markovic\orcid{0000-0001-6764-073X}\inst{\ref{aff55}}
\and M.~Martinelli\orcid{0000-0002-6943-7732}\inst{\ref{aff32},\ref{aff71}}
\and N.~Martinet\orcid{0000-0003-2786-7790}\inst{\ref{aff25}}
\and F.~Marulli\orcid{0000-0002-8850-0303}\inst{\ref{aff72},\ref{aff16},\ref{aff17}}
\and R.~Massey\orcid{0000-0002-6085-3780}\inst{\ref{aff73}}
\and S.~Maurogordato\inst{\ref{aff74}}
\and E.~Medinaceli\orcid{0000-0002-4040-7783}\inst{\ref{aff16}}
\and M.~Melchior\inst{\ref{aff75}}
\and Y.~Mellier\inst{\ref{aff76},\ref{aff77}}
\and M.~Meneghetti\orcid{0000-0003-1225-7084}\inst{\ref{aff16},\ref{aff17}}
\and E.~Merlin\orcid{0000-0001-6870-8900}\inst{\ref{aff32}}
\and G.~Meylan\inst{\ref{aff40}}
\and L.~Moscardini\orcid{0000-0002-3473-6716}\inst{\ref{aff72},\ref{aff16},\ref{aff17}}
\and E.~Munari\orcid{0000-0002-1751-5946}\inst{\ref{aff1},\ref{aff3}}
\and S.-M.~Niemi\inst{\ref{aff78}}
\and C.~Padilla\orcid{0000-0001-7951-0166}\inst{\ref{aff79}}
\and S.~Paltani\orcid{0000-0002-8108-9179}\inst{\ref{aff44}}
\and F.~Pasian\orcid{0000-0002-4869-3227}\inst{\ref{aff1}}
\and K.~Pedersen\inst{\ref{aff80}}
\and W.~J.~Percival\orcid{0000-0002-0644-5727}\inst{\ref{aff81},\ref{aff82},\ref{aff83}}
\and V.~Pettorino\inst{\ref{aff78}}
\and S.~Pires\orcid{0000-0002-0249-2104}\inst{\ref{aff47}}
\and G.~Polenta\orcid{0000-0003-4067-9196}\inst{\ref{aff84}}
\and M.~Poncet\inst{\ref{aff85}}
\and L.~A.~Popa\inst{\ref{aff86}}
\and L.~Pozzetti\orcid{0000-0001-7085-0412}\inst{\ref{aff16}}
\and F.~Raison\orcid{0000-0002-7819-6918}\inst{\ref{aff18}}
\and A.~Renzi\orcid{0000-0001-9856-1970}\inst{\ref{aff87},\ref{aff46}}
\and G.~Riccio\inst{\ref{aff23}}
\and E.~Romelli\orcid{0000-0003-3069-9222}\inst{\ref{aff1}}
\and M.~Roncarelli\orcid{0000-0001-9587-7822}\inst{\ref{aff16}}
\and R.~Saglia\orcid{0000-0003-0378-7032}\inst{\ref{aff8},\ref{aff18}}
\and Z.~Sakr\orcid{0000-0002-4823-3757}\inst{\ref{aff12},\ref{aff60},\ref{aff88}}
\and J.-C.~Salvignol\inst{\ref{aff78}}
\and A.~G.~S\'anchez\orcid{0000-0003-1198-831X}\inst{\ref{aff18}}
\and D.~Sapone\orcid{0000-0001-7089-4503}\inst{\ref{aff89}}
\and B.~Sartoris\orcid{0000-0003-1337-5269}\inst{\ref{aff8},\ref{aff1}}
\and M.~Schirmer\orcid{0000-0003-2568-9994}\inst{\ref{aff61}}
\and A.~Secroun\orcid{0000-0003-0505-3710}\inst{\ref{aff65}}
\and S.~Serrano\orcid{0000-0002-0211-2861}\inst{\ref{aff48},\ref{aff90},\ref{aff49}}
\and C.~Sirignano\orcid{0000-0002-0995-7146}\inst{\ref{aff87},\ref{aff46}}
\and G.~Sirri\orcid{0000-0003-2626-2853}\inst{\ref{aff17}}
\and L.~Stanco\orcid{0000-0002-9706-5104}\inst{\ref{aff46}}
\and J.~Steinwagner\orcid{0000-0001-7443-1047}\inst{\ref{aff18}}
\and P.~Tallada-Cresp\'{i}\orcid{0000-0002-1336-8328}\inst{\ref{aff29},\ref{aff30}}
\and A.~N.~Taylor\inst{\ref{aff35}}
\and I.~Tereno\inst{\ref{aff42},\ref{aff91}}
\and R.~Toledo-Moreo\orcid{0000-0002-2997-4859}\inst{\ref{aff92}}
\and F.~Torradeflot\orcid{0000-0003-1160-1517}\inst{\ref{aff30},\ref{aff29}}
\and I.~Tutusaus\orcid{0000-0002-3199-0399}\inst{\ref{aff60}}
\and L.~Valenziano\orcid{0000-0002-1170-0104}\inst{\ref{aff16},\ref{aff93}}
\and T.~Vassallo\orcid{0000-0001-6512-6358}\inst{\ref{aff8},\ref{aff1}}
\and G.~Verdoes~Kleijn\orcid{0000-0001-5803-2580}\inst{\ref{aff94}}
\and Y.~Wang\orcid{0000-0002-4749-2984}\inst{\ref{aff95}}
\and J.~Weller\orcid{0000-0002-8282-2010}\inst{\ref{aff8},\ref{aff18}}
\and A.~Zacchei\orcid{0000-0003-0396-1192}\inst{\ref{aff1},\ref{aff3}}
\and G.~Zamorani\orcid{0000-0002-2318-301X}\inst{\ref{aff16}}
\and E.~Zucca\orcid{0000-0002-5845-8132}\inst{\ref{aff16}}
\and A.~Biviano\orcid{0000-0002-0857-0732}\inst{\ref{aff1},\ref{aff3}}
\and M.~Bolzonella\orcid{0000-0003-3278-4607}\inst{\ref{aff16}}
\and E.~Bozzo\orcid{0000-0002-8201-1525}\inst{\ref{aff44}}
\and C.~Burigana\orcid{0000-0002-3005-5796}\inst{\ref{aff96},\ref{aff93}}
\and M.~Calabrese\orcid{0000-0002-2637-2422}\inst{\ref{aff97},\ref{aff28}}
\and D.~Di~Ferdinando\inst{\ref{aff17}}
\and J.~A.~Escartin~Vigo\inst{\ref{aff18}}
\and F.~Finelli\orcid{0000-0002-6694-3269}\inst{\ref{aff16},\ref{aff93}}
\and J.~Gracia-Carpio\inst{\ref{aff18}}
\and S.~Matthew\orcid{0000-0001-8448-1697}\inst{\ref{aff35}}
\and N.~Mauri\orcid{0000-0001-8196-1548}\inst{\ref{aff33},\ref{aff17}}
\and A.~Pezzotta\orcid{0000-0003-0726-2268}\inst{\ref{aff18}}
\and M.~P\"ontinen\orcid{0000-0001-5442-2530}\inst{\ref{aff67}}
\and C.~Porciani\orcid{0000-0002-7797-2508}\inst{\ref{aff70}}
\and V.~Scottez\inst{\ref{aff76},\ref{aff98}}
\and M.~Tenti\orcid{0000-0002-4254-5901}\inst{\ref{aff17}}
\and M.~Viel\orcid{0000-0002-2642-5707}\inst{\ref{aff3},\ref{aff1},\ref{aff14},\ref{aff2},\ref{aff4}}
\and M.~Wiesmann\orcid{0009-0000-8199-5860}\inst{\ref{aff54}}
\and Y.~Akrami\orcid{0000-0002-2407-7956}\inst{\ref{aff99},\ref{aff100}}
\and V.~Allevato\orcid{0000-0001-7232-5152}\inst{\ref{aff23}}
\and S.~Anselmi\orcid{0000-0002-3579-9583}\inst{\ref{aff46},\ref{aff87},\ref{aff101}}
\and M.~Archidiacono\orcid{0000-0003-4952-9012}\inst{\ref{aff53},\ref{aff102}}
\and F.~Atrio-Barandela\orcid{0000-0002-2130-2513}\inst{\ref{aff103}}
\and A.~Balaguera-Antolinez\orcid{0000-0001-5028-3035}\inst{\ref{aff34},\ref{aff104}}
\and M.~Ballardini\orcid{0000-0003-4481-3559}\inst{\ref{aff105},\ref{aff16},\ref{aff106}}
\and D.~Bertacca\orcid{0000-0002-2490-7139}\inst{\ref{aff87},\ref{aff52},\ref{aff46}}
\and M.~Bethermin\orcid{0000-0002-3915-2015}\inst{\ref{aff107},\ref{aff25}}
\and A.~Blanchard\orcid{0000-0001-8555-9003}\inst{\ref{aff60}}
\and L.~Blot\orcid{0000-0002-9622-7167}\inst{\ref{aff108},\ref{aff101}}
\and H.~B\"ohringer\orcid{0000-0001-8241-4204}\inst{\ref{aff18},\ref{aff5},\ref{aff109}}
\and S.~Bruton\orcid{0000-0002-6503-5218}\inst{\ref{aff110}}
\and R.~Cabanac\orcid{0000-0001-6679-2600}\inst{\ref{aff60}}
\and A.~Calabro\orcid{0000-0003-2536-1614}\inst{\ref{aff32}}
\and G.~Ca\~nas-Herrera\orcid{0000-0003-2796-2149}\inst{\ref{aff78},\ref{aff111}}
\and A.~Cappi\inst{\ref{aff16},\ref{aff74}}
\and F.~Caro\inst{\ref{aff32}}
\and C.~S.~Carvalho\inst{\ref{aff91}}
\and K.~C.~Chambers\orcid{0000-0001-6965-7789}\inst{\ref{aff112}}
\and A.~R.~Cooray\orcid{0000-0002-3892-0190}\inst{\ref{aff113}}
\and B.~De~Caro\inst{\ref{aff28}}
\and S.~de~la~Torre\inst{\ref{aff25}}
\and G.~Desprez\orcid{0000-0001-8325-1742}\inst{\ref{aff114}}
\and A.~D\'iaz-S\'anchez\orcid{0000-0003-0748-4768}\inst{\ref{aff115}}
\and J.~J.~Diaz\inst{\ref{aff116}}
\and S.~Di~Domizio\orcid{0000-0003-2863-5895}\inst{\ref{aff20},\ref{aff21}}
\and H.~Dole\orcid{0000-0002-9767-3839}\inst{\ref{aff11}}
\and S.~Escoffier\orcid{0000-0002-2847-7498}\inst{\ref{aff65}}
\and A.~G.~Ferrari\orcid{0009-0005-5266-4110}\inst{\ref{aff33},\ref{aff17}}
\and P.~G.~Ferreira\orcid{0000-0002-3021-2851}\inst{\ref{aff117}}
\and I.~Ferrero\orcid{0000-0002-1295-1132}\inst{\ref{aff54}}
\and A.~Finoguenov\orcid{0000-0002-4606-5403}\inst{\ref{aff67}}
\and A.~Fontana\orcid{0000-0003-3820-2823}\inst{\ref{aff32}}
\and F.~Fornari\orcid{0000-0003-2979-6738}\inst{\ref{aff93}}
\and L.~Gabarra\orcid{0000-0002-8486-8856}\inst{\ref{aff117}}
\and K.~Ganga\orcid{0000-0001-8159-8208}\inst{\ref{aff118}}
\and J.~Garc\'ia-Bellido\orcid{0000-0002-9370-8360}\inst{\ref{aff99}}
\and T.~Gasparetto\orcid{0000-0002-7913-4866}\inst{\ref{aff1}}
\and V.~Gautard\inst{\ref{aff119}}
\and E.~Gaztanaga\orcid{0000-0001-9632-0815}\inst{\ref{aff49},\ref{aff48},\ref{aff120}}
\and F.~Giacomini\orcid{0000-0002-3129-2814}\inst{\ref{aff17}}
\and F.~Gianotti\orcid{0000-0003-4666-119X}\inst{\ref{aff16}}
\and G.~Gozaliasl\orcid{0000-0002-0236-919X}\inst{\ref{aff121},\ref{aff67}}
\and C.~M.~Gutierrez\orcid{0000-0001-7854-783X}\inst{\ref{aff122}}
\and A.~Hall\orcid{0000-0002-3139-8651}\inst{\ref{aff35}}
\and H.~Hildebrandt\orcid{0000-0002-9814-3338}\inst{\ref{aff123}}
\and J.~Hjorth\orcid{0000-0002-4571-2306}\inst{\ref{aff80}}
\and A.~Jimenez~Mu\~noz\orcid{0009-0004-5252-185X}\inst{\ref{aff124}}
\and J.~J.~E.~Kajava\orcid{0000-0002-3010-8333}\inst{\ref{aff125},\ref{aff126}}
\and V.~Kansal\orcid{0000-0002-4008-6078}\inst{\ref{aff127},\ref{aff128}}
\and D.~Karagiannis\orcid{0000-0002-4927-0816}\inst{\ref{aff129},\ref{aff130}}
\and C.~C.~Kirkpatrick\inst{\ref{aff64}}
\and A.~M.~C.~Le~Brun\orcid{0000-0002-0936-4594}\inst{\ref{aff101}}
\and J.~Le~Graet\orcid{0000-0001-6523-7971}\inst{\ref{aff65}}
\and L.~Legrand\orcid{0000-0003-0610-5252}\inst{\ref{aff131}}
\and J.~Lesgourgues\orcid{0000-0001-7627-353X}\inst{\ref{aff31}}
\and T.~I.~Liaudat\orcid{0000-0002-9104-314X}\inst{\ref{aff132}}
\and A.~Loureiro\orcid{0000-0002-4371-0876}\inst{\ref{aff133},\ref{aff134}}
\and G.~Maggio\orcid{0000-0003-4020-4836}\inst{\ref{aff1}}
\and M.~Magliocchetti\orcid{0000-0001-9158-4838}\inst{\ref{aff45}}
\and F.~Mannucci\orcid{0000-0002-4803-2381}\inst{\ref{aff135}}
\and R.~Maoli\orcid{0000-0002-6065-3025}\inst{\ref{aff136},\ref{aff32}}
\and C.~J.~A.~P.~Martins\orcid{0000-0002-4886-9261}\inst{\ref{aff137},\ref{aff138}}
\and L.~Maurin\orcid{0000-0002-8406-0857}\inst{\ref{aff11}}
\and R.~B.~Metcalf\orcid{0000-0003-3167-2574}\inst{\ref{aff72},\ref{aff16}}
\and M.~Miluzio\inst{\ref{aff38},\ref{aff139}}
\and A.~Montoro\orcid{0000-0003-4730-8590}\inst{\ref{aff49},\ref{aff48}}
\and A.~Mora\orcid{0000-0002-1922-8529}\inst{\ref{aff140}}
\and C.~Moretti\orcid{0000-0003-3314-8936}\inst{\ref{aff14},\ref{aff4},\ref{aff1},\ref{aff3},\ref{aff2}}
\and G.~Morgante\inst{\ref{aff16}}
\and S.~Nadathur\orcid{0000-0001-9070-3102}\inst{\ref{aff120}}
\and Nicholas~A.~Walton\orcid{0000-0003-3983-8778}\inst{\ref{aff141}}
\and L.~Pagano\orcid{0000-0003-1820-5998}\inst{\ref{aff105},\ref{aff106}}
\and L.~Patrizii\inst{\ref{aff17}}
\and V.~Popa\orcid{0000-0002-9118-8330}\inst{\ref{aff86}}
\and D.~Potter\orcid{0000-0002-0757-5195}\inst{\ref{aff10}}
\and I.~Risso\orcid{0000-0003-2525-7761}\inst{\ref{aff142}}
\and P.-F.~Rocci\inst{\ref{aff11}}
\and M.~Sahl\'en\orcid{0000-0003-0973-4804}\inst{\ref{aff143}}
\and E.~Sarpa\orcid{0000-0002-1256-655X}\inst{\ref{aff14},\ref{aff4},\ref{aff2}}
\and A.~Schneider\orcid{0000-0001-7055-8104}\inst{\ref{aff10}}
\and M.~Sereno\orcid{0000-0003-0302-0325}\inst{\ref{aff16},\ref{aff17}}
\and A.~Spurio~Mancini\orcid{0000-0001-5698-0990}\inst{\ref{aff144},\ref{aff145}}
\and J.~Stadel\orcid{0000-0001-7565-8622}\inst{\ref{aff10}}
\and K.~Tanidis\inst{\ref{aff117}}
\and C.~Tao\orcid{0000-0001-7961-8177}\inst{\ref{aff65}}
\and N.~Tessore\orcid{0000-0002-9696-7931}\inst{\ref{aff63}}
\and G.~Testera\inst{\ref{aff21}}
\and R.~Teyssier\orcid{0000-0001-7689-0933}\inst{\ref{aff146}}
\and S.~Toft\orcid{0000-0003-3631-7176}\inst{\ref{aff147},\ref{aff148}}
\and S.~Tosi\orcid{0000-0002-7275-9193}\inst{\ref{aff20},\ref{aff21}}
\and A.~Troja\orcid{0000-0003-0239-4595}\inst{\ref{aff87},\ref{aff46}}
\and M.~Tucci\inst{\ref{aff44}}
\and C.~Valieri\inst{\ref{aff17}}
\and J.~Valiviita\orcid{0000-0001-6225-3693}\inst{\ref{aff67},\ref{aff68}}
\and D.~Vergani\orcid{0000-0003-0898-2216}\inst{\ref{aff16}}
\and G.~Verza\orcid{0000-0002-1886-8348}\inst{\ref{aff149},\ref{aff150}}
\and P.~Vielzeuf\orcid{0000-0003-2035-9339}\inst{\ref{aff65}}}
										   
\institute{INAF-Osservatorio Astronomico di Trieste, Via G. B. Tiepolo 11, 34143 Trieste, Italy\label{aff1}
\and
INFN, Sezione di Trieste, Via Valerio 2, 34127 Trieste TS, Italy\label{aff2}
\and
IFPU, Institute for Fundamental Physics of the Universe, via Beirut 2, 34151 Trieste, Italy\label{aff3}
\and
ICSC - Centro Nazionale di Ricerca in High Performance Computing, Big Data e Quantum Computing, Via Magnanelli 2, Bologna, Italy\label{aff4}
\and
Ludwig-Maximilians-University, Schellingstrasse 4, 80799 Munich, Germany\label{aff5}
\and
Donostia International Physics Center (DIPC), Paseo Manuel de Lardizabal, 4, 20018, Donostia-San Sebasti\'an, Guipuzkoa, Spain\label{aff6}
\and
IKERBASQUE, Basque Foundation for Science, 48013, Bilbao, Spain\label{aff7}
\and
Universit\"ats-Sternwarte M\"unchen, Fakult\"at f\"ur Physik, Ludwig-Maximilians-Universit\"at M\"unchen, Scheinerstrasse 1, 81679 M\"unchen, Germany\label{aff8}
\and
Dipartimento di Fisica - Sezione di Astronomia, Universit\`a di Trieste, Via Tiepolo 11, 34131 Trieste, Italy\label{aff9}
\and
Department of Astrophysics, University of Zurich, Winterthurerstrasse 190, 8057 Zurich, Switzerland\label{aff10}
\and
Universit\'e Paris-Saclay, CNRS, Institut d'astrophysique spatiale, 91405, Orsay, France\label{aff11}
\and
Institut f\"ur Theoretische Physik, University of Heidelberg, Philosophenweg 16, 69120 Heidelberg, Germany\label{aff12}
\and
INAF-Osservatorio Astronomico di Brera, Via Brera 28, 20122 Milano, Italy\label{aff13}
\and
SISSA, International School for Advanced Studies, Via Bonomea 265, 34136 Trieste TS, Italy\label{aff14}
\and
Dipartimento di Fisica e Astronomia, Universit\`a di Bologna, Via Gobetti 93/2, 40129 Bologna, Italy\label{aff15}
\and
INAF-Osservatorio di Astrofisica e Scienza dello Spazio di Bologna, Via Piero Gobetti 93/3, 40129 Bologna, Italy\label{aff16}
\and
INFN-Sezione di Bologna, Viale Berti Pichat 6/2, 40127 Bologna, Italy\label{aff17}
\and
Max Planck Institute for Extraterrestrial Physics, Giessenbachstr. 1, 85748 Garching, Germany\label{aff18}
\and
INAF-Osservatorio Astrofisico di Torino, Via Osservatorio 20, 10025 Pino Torinese (TO), Italy\label{aff19}
\and
Dipartimento di Fisica, Universit\`a di Genova, Via Dodecaneso 33, 16146, Genova, Italy\label{aff20}
\and
INFN-Sezione di Genova, Via Dodecaneso 33, 16146, Genova, Italy\label{aff21}
\and
Department of Physics "E. Pancini", University Federico II, Via Cinthia 6, 80126, Napoli, Italy\label{aff22}
\and
INAF-Osservatorio Astronomico di Capodimonte, Via Moiariello 16, 80131 Napoli, Italy\label{aff23}
\and
INFN section of Naples, Via Cinthia 6, 80126, Napoli, Italy\label{aff24}
\and
Aix-Marseille Universit\'e, CNRS, CNES, LAM, Marseille, France\label{aff25}
\and
Dipartimento di Fisica, Universit\`a degli Studi di Torino, Via P. Giuria 1, 10125 Torino, Italy\label{aff26}
\and
INFN-Sezione di Torino, Via P. Giuria 1, 10125 Torino, Italy\label{aff27}
\and
INAF-IASF Milano, Via Alfonso Corti 12, 20133 Milano, Italy\label{aff28}
\and
Centro de Investigaciones Energ\'eticas, Medioambientales y Tecnol\'ogicas (CIEMAT), Avenida Complutense 40, 28040 Madrid, Spain\label{aff29}
\and
Port d'Informaci\'{o} Cient\'{i}fica, Campus UAB, C. Albareda s/n, 08193 Bellaterra (Barcelona), Spain\label{aff30}
\and
Institute for Theoretical Particle Physics and Cosmology (TTK), RWTH Aachen University, 52056 Aachen, Germany\label{aff31}
\and
INAF-Osservatorio Astronomico di Roma, Via Frascati 33, 00078 Monteporzio Catone, Italy\label{aff32}
\and
Dipartimento di Fisica e Astronomia "Augusto Righi" - Alma Mater Studiorum Universit\`a di Bologna, Viale Berti Pichat 6/2, 40127 Bologna, Italy\label{aff33}
\and
Instituto de Astrof\'isica de Canarias, Calle V\'ia L\'actea s/n, 38204, San Crist\'obal de La Laguna, Tenerife, Spain\label{aff34}
\and
Institute for Astronomy, University of Edinburgh, Royal Observatory, Blackford Hill, Edinburgh EH9 3HJ, UK\label{aff35}
\and
Jodrell Bank Centre for Astrophysics, Department of Physics and Astronomy, University of Manchester, Oxford Road, Manchester M13 9PL, UK\label{aff36}
\and
European Space Agency/ESRIN, Largo Galileo Galilei 1, 00044 Frascati, Roma, Italy\label{aff37}
\and
ESAC/ESA, Camino Bajo del Castillo, s/n., Urb. Villafranca del Castillo, 28692 Villanueva de la Ca\~nada, Madrid, Spain\label{aff38}
\and
Universit\'e Claude Bernard Lyon 1, CNRS/IN2P3, IP2I Lyon, UMR 5822, Villeurbanne, F-69100, France\label{aff39}
\and
Institute of Physics, Laboratory of Astrophysics, Ecole Polytechnique F\'ed\'erale de Lausanne (EPFL), Observatoire de Sauverny, 1290 Versoix, Switzerland\label{aff40}
\and
UCB Lyon 1, CNRS/IN2P3, IUF, IP2I Lyon, 4 rue Enrico Fermi, 69622 Villeurbanne, France\label{aff41}
\and
Departamento de F\'isica, Faculdade de Ci\^encias, Universidade de Lisboa, Edif\'icio C8, Campo Grande, PT1749-016 Lisboa, Portugal\label{aff42}
\and
Instituto de Astrof\'isica e Ci\^encias do Espa\c{c}o, Faculdade de Ci\^encias, Universidade de Lisboa, Campo Grande, 1749-016 Lisboa, Portugal\label{aff43}
\and
Department of Astronomy, University of Geneva, ch. d'Ecogia 16, 1290 Versoix, Switzerland\label{aff44}
\and
INAF-Istituto di Astrofisica e Planetologia Spaziali, via del Fosso del Cavaliere, 100, 00100 Roma, Italy\label{aff45}
\and
INFN-Padova, Via Marzolo 8, 35131 Padova, Italy\label{aff46}
\and
Universit\'e Paris-Saclay, Universit\'e Paris Cit\'e, CEA, CNRS, AIM, 91191, Gif-sur-Yvette, France\label{aff47}
\and
Institut d'Estudis Espacials de Catalunya (IEEC),  Edifici RDIT, Campus UPC, 08860 Castelldefels, Barcelona, Spain\label{aff48}
\and
Institute of Space Sciences (ICE, CSIC), Campus UAB, Carrer de Can Magrans, s/n, 08193 Barcelona, Spain\label{aff49}
\and
Istituto Nazionale di Fisica Nucleare, Sezione di Bologna, Via Irnerio 46, 40126 Bologna, Italy\label{aff50}
\and
FRACTAL S.L.N.E., calle Tulip\'an 2, Portal 13 1A, 28231, Las Rozas de Madrid, Spain\label{aff51}
\and
INAF-Osservatorio Astronomico di Padova, Via dell'Osservatorio 5, 35122 Padova, Italy\label{aff52}
\and
Dipartimento di Fisica "Aldo Pontremoli", Universit\`a degli Studi di Milano, Via Celoria 16, 20133 Milano, Italy\label{aff53}
\and
Institute of Theoretical Astrophysics, University of Oslo, P.O. Box 1029 Blindern, 0315 Oslo, Norway\label{aff54}
\and
Jet Propulsion Laboratory, California Institute of Technology, 4800 Oak Grove Drive, Pasadena, CA, 91109, USA\label{aff55}
\and
Felix Hormuth Engineering, Goethestr. 17, 69181 Leimen, Germany\label{aff56}
\and
Technical University of Denmark, Elektrovej 327, 2800 Kgs. Lyngby, Denmark\label{aff57}
\and
Cosmic Dawn Center (DAWN), Denmark\label{aff58}
\and
Universit\'e Paris-Saclay, CNRS/IN2P3, IJCLab, 91405 Orsay, France\label{aff59}
\and
Institut de Recherche en Astrophysique et Plan\'etologie (IRAP), Universit\'e de Toulouse, CNRS, UPS, CNES, 14 Av. Edouard Belin, 31400 Toulouse, France\label{aff60}
\and
Max-Planck-Institut f\"ur Astronomie, K\"onigstuhl 17, 69117 Heidelberg, Germany\label{aff61}
\and
NASA Goddard Space Flight Center, Greenbelt, MD 20771, USA\label{aff62}
\and
Department of Physics and Astronomy, University College London, Gower Street, London WC1E 6BT, UK\label{aff63}
\and
Department of Physics and Helsinki Institute of Physics, Gustaf H\"allstr\"omin katu 2, 00014 University of Helsinki, Finland\label{aff64}
\and
Aix-Marseille Universit\'e, CNRS/IN2P3, CPPM, Marseille, France\label{aff65}
\and
Universit\'e de Gen\`eve, D\'epartement de Physique Th\'eorique and Centre for Astroparticle Physics, 24 quai Ernest-Ansermet, CH-1211 Gen\`eve 4, Switzerland\label{aff66}
\and
Department of Physics, P.O. Box 64, 00014 University of Helsinki, Finland\label{aff67}
\and
Helsinki Institute of Physics, Gustaf H{\"a}llstr{\"o}min katu 2, University of Helsinki, Helsinki, Finland\label{aff68}
\and
NOVA optical infrared instrumentation group at ASTRON, Oude Hoogeveensedijk 4, 7991PD, Dwingeloo, The Netherlands\label{aff69}
\and
Universit\"at Bonn, Argelander-Institut f\"ur Astronomie, Auf dem H\"ugel 71, 53121 Bonn, Germany\label{aff70}
\and
INFN-Sezione di Roma, Piazzale Aldo Moro, 2 - c/o Dipartimento di Fisica, Edificio G. Marconi, 00185 Roma, Italy\label{aff71}
\and
Dipartimento di Fisica e Astronomia "Augusto Righi" - Alma Mater Studiorum Universit\`a di Bologna, via Piero Gobetti 93/2, 40129 Bologna, Italy\label{aff72}
\and
Department of Physics, Institute for Computational Cosmology, Durham University, South Road, DH1 3LE, UK\label{aff73}
\and
Universit\'e C\^{o}te d'Azur, Observatoire de la C\^{o}te d'Azur, CNRS, Laboratoire Lagrange, Bd de l'Observatoire, CS 34229, 06304 Nice cedex 4, France\label{aff74}
\and
University of Applied Sciences and Arts of Northwestern Switzerland, School of Engineering, 5210 Windisch, Switzerland\label{aff75}
\and
Institut d'Astrophysique de Paris, 98bis Boulevard Arago, 75014, Paris, France\label{aff76}
\and
Institut d'Astrophysique de Paris, UMR 7095, CNRS, and Sorbonne Universit\'e, 98 bis boulevard Arago, 75014 Paris, France\label{aff77}
\and
European Space Agency/ESTEC, Keplerlaan 1, 2201 AZ Noordwijk, The Netherlands\label{aff78}
\and
Institut de F\'{i}sica d'Altes Energies (IFAE), The Barcelona Institute of Science and Technology, Campus UAB, 08193 Bellaterra (Barcelona), Spain\label{aff79}
\and
DARK, Niels Bohr Institute, University of Copenhagen, Jagtvej 155, 2200 Copenhagen, Denmark\label{aff80}
\and
Waterloo Centre for Astrophysics, University of Waterloo, Waterloo, Ontario N2L 3G1, Canada\label{aff81}
\and
Department of Physics and Astronomy, University of Waterloo, Waterloo, Ontario N2L 3G1, Canada\label{aff82}
\and
Perimeter Institute for Theoretical Physics, Waterloo, Ontario N2L 2Y5, Canada\label{aff83}
\and
Space Science Data Center, Italian Space Agency, via del Politecnico snc, 00133 Roma, Italy\label{aff84}
\and
Centre National d'Etudes Spatiales -- Centre spatial de Toulouse, 18 avenue Edouard Belin, 31401 Toulouse Cedex 9, France\label{aff85}
\and
Institute of Space Science, Str. Atomistilor, nr. 409 M\u{a}gurele, Ilfov, 077125, Romania\label{aff86}
\and
Dipartimento di Fisica e Astronomia "G. Galilei", Universit\`a di Padova, Via Marzolo 8, 35131 Padova, Italy\label{aff87}
\and
Universit\'e St Joseph; Faculty of Sciences, Beirut, Lebanon\label{aff88}
\and
Departamento de F\'isica, FCFM, Universidad de Chile, Blanco Encalada 2008, Santiago, Chile\label{aff89}
\and
Satlantis, University Science Park, Sede Bld 48940, Leioa-Bilbao, Spain\label{aff90}
\and
Instituto de Astrof\'isica e Ci\^encias do Espa\c{c}o, Faculdade de Ci\^encias, Universidade de Lisboa, Tapada da Ajuda, 1349-018 Lisboa, Portugal\label{aff91}
\and
Universidad Polit\'ecnica de Cartagena, Departamento de Electr\'onica y Tecnolog\'ia de Computadoras,  Plaza del Hospital 1, 30202 Cartagena, Spain\label{aff92}
\and
INFN-Bologna, Via Irnerio 46, 40126 Bologna, Italy\label{aff93}
\and
Kapteyn Astronomical Institute, University of Groningen, PO Box 800, 9700 AV Groningen, The Netherlands\label{aff94}
\and
Infrared Processing and Analysis Center, California Institute of Technology, Pasadena, CA 91125, USA\label{aff95}
\and
INAF, Istituto di Radioastronomia, Via Piero Gobetti 101, 40129 Bologna, Italy\label{aff96}
\and
Astronomical Observatory of the Autonomous Region of the Aosta Valley (OAVdA), Loc. Lignan 39, I-11020, Nus (Aosta Valley), Italy\label{aff97}
\and
Junia, EPA department, 41 Bd Vauban, 59800 Lille, France\label{aff98}
\and
Instituto de F\'isica Te\'orica UAM-CSIC, Campus de Cantoblanco, 28049 Madrid, Spain\label{aff99}
\and
CERCA/ISO, Department of Physics, Case Western Reserve University, 10900 Euclid Avenue, Cleveland, OH 44106, USA\label{aff100}
\and
Laboratoire Univers et Th\'eorie, Observatoire de Paris, Universit\'e PSL, Universit\'e Paris Cit\'e, CNRS, 92190 Meudon, France\label{aff101}
\and
INFN-Sezione di Milano, Via Celoria 16, 20133 Milano, Italy\label{aff102}
\and
Departamento de F{\'\i}sica Fundamental. Universidad de Salamanca. Plaza de la Merced s/n. 37008 Salamanca, Spain\label{aff103}
\and
Departamento de Astrof\'isica, Universidad de La Laguna, 38206, La Laguna, Tenerife, Spain\label{aff104}
\and
Dipartimento di Fisica e Scienze della Terra, Universit\`a degli Studi di Ferrara, Via Giuseppe Saragat 1, 44122 Ferrara, Italy\label{aff105}
\and
Istituto Nazionale di Fisica Nucleare, Sezione di Ferrara, Via Giuseppe Saragat 1, 44122 Ferrara, Italy\label{aff106}
\and
Universit\'e de Strasbourg, CNRS, Observatoire astronomique de Strasbourg, UMR 7550, 67000 Strasbourg, France\label{aff107}
\and
Center for Data-Driven Discovery, Kavli IPMU (WPI), UTIAS, The University of Tokyo, Kashiwa, Chiba 277-8583, Japan\label{aff108}
\and
Max-Planck-Institut f\"ur Physik, Boltzmannstr. 8, 85748 Garching, Germany\label{aff109}
\and
Minnesota Institute for Astrophysics, University of Minnesota, 116 Church St SE, Minneapolis, MN 55455, USA\label{aff110}
\and
Institute Lorentz, Leiden University, Niels Bohrweg 2, 2333 CA Leiden, The Netherlands\label{aff111}
\and
Institute for Astronomy, University of Hawaii, 2680 Woodlawn Drive, Honolulu, HI 96822, USA\label{aff112}
\and
Department of Physics \& Astronomy, University of California Irvine, Irvine CA 92697, USA\label{aff113}
\and
Department of Astronomy \& Physics and Institute for Computational Astrophysics, Saint Mary's University, 923 Robie Street, Halifax, Nova Scotia, B3H 3C3, Canada\label{aff114}
\and
Departamento F\'isica Aplicada, Universidad Polit\'ecnica de Cartagena, Campus Muralla del Mar, 30202 Cartagena, Murcia, Spain\label{aff115}
\and
Instituto de Astrof\'isica de Canarias (IAC); Departamento de Astrof\'isica, Universidad de La Laguna (ULL), 38200, La Laguna, Tenerife, Spain\label{aff116}
\and
Department of Physics, Oxford University, Keble Road, Oxford OX1 3RH, UK\label{aff117}
\and
Universit\'e Paris Cit\'e, CNRS, Astroparticule et Cosmologie, 75013 Paris, France\label{aff118}
\and
CEA Saclay, DFR/IRFU, Service d'Astrophysique, Bat. 709, 91191 Gif-sur-Yvette, France\label{aff119}
\and
Institute of Cosmology and Gravitation, University of Portsmouth, Portsmouth PO1 3FX, UK\label{aff120}
\and
Department of Computer Science, Aalto University, PO Box 15400, Espoo, FI-00 076, Finland\label{aff121}
\and
Instituto de Astrof\'\i sica de Canarias, c/ Via Lactea s/n, La Laguna E-38200, Spain. Departamento de Astrof\'\i sica de la Universidad de La Laguna, Avda. Francisco Sanchez, La Laguna, E-38200, Spain\label{aff122}
\and
Ruhr University Bochum, Faculty of Physics and Astronomy, Astronomical Institute (AIRUB), German Centre for Cosmological Lensing (GCCL), 44780 Bochum, Germany\label{aff123}
\and
Univ. Grenoble Alpes, CNRS, Grenoble INP, LPSC-IN2P3, 53, Avenue des Martyrs, 38000, Grenoble, France\label{aff124}
\and
Department of Physics and Astronomy, Vesilinnantie 5, 20014 University of Turku, Finland\label{aff125}
\and
Serco for European Space Agency (ESA), Camino bajo del Castillo, s/n, Urbanizacion Villafranca del Castillo, Villanueva de la Ca\~nada, 28692 Madrid, Spain\label{aff126}
\and
ARC Centre of Excellence for Dark Matter Particle Physics, Melbourne, Australia\label{aff127}
\and
Centre for Astrophysics \& Supercomputing, Swinburne University of Technology,  Hawthorn, Victoria 3122, Australia\label{aff128}
\and
School of Physics and Astronomy, Queen Mary University of London, Mile End Road, London E1 4NS, UK\label{aff129}
\and
Department of Physics and Astronomy, University of the Western Cape, Bellville, Cape Town, 7535, South Africa\label{aff130}
\and
ICTP South American Institute for Fundamental Research, Instituto de F\'{\i}sica Te\'orica, Universidade Estadual Paulista, S\~ao Paulo, Brazil\label{aff131}
\and
IRFU, CEA, Universit\'e Paris-Saclay 91191 Gif-sur-Yvette Cedex, France\label{aff132}
\and
Oskar Klein Centre for Cosmoparticle Physics, Department of Physics, Stockholm University, Stockholm, SE-106 91, Sweden\label{aff133}
\and
Astrophysics Group, Blackett Laboratory, Imperial College London, London SW7 2AZ, UK\label{aff134}
\and
INAF-Osservatorio Astrofisico di Arcetri, Largo E. Fermi 5, 50125, Firenze, Italy\label{aff135}
\and
Dipartimento di Fisica, Sapienza Universit\`a di Roma, Piazzale Aldo Moro 2, 00185 Roma, Italy\label{aff136}
\and
Centro de Astrof\'{\i}sica da Universidade do Porto, Rua das Estrelas, 4150-762 Porto, Portugal\label{aff137}
\and
Instituto de Astrof\'isica e Ci\^encias do Espa\c{c}o, Universidade do Porto, CAUP, Rua das Estrelas, PT4150-762 Porto, Portugal\label{aff138}
\and
HE Space for European Space Agency (ESA), Camino bajo del Castillo, s/n, Urbanizacion Villafranca del Castillo, Villanueva de la Ca\~nada, 28692 Madrid, Spain\label{aff139}
\and
Aurora Technology for European Space Agency (ESA), Camino bajo del Castillo, s/n, Urbanizacion Villafranca del Castillo, Villanueva de la Ca\~nada, 28692 Madrid, Spain\label{aff140}
\and
Institute of Astronomy, University of Cambridge, Madingley Road, Cambridge CB3 0HA, UK\label{aff141}
\and
Dipartimento di Fisica, Universit\`a degli studi di Genova, and INFN-Sezione di Genova, via Dodecaneso 33, 16146, Genova, Italy\label{aff142}
\and
Theoretical astrophysics, Department of Physics and Astronomy, Uppsala University, Box 515, 751 20 Uppsala, Sweden\label{aff143}
\and
Department of Physics, Royal Holloway, University of London, TW20 0EX, UK\label{aff144}
\and
Mullard Space Science Laboratory, University College London, Holmbury St Mary, Dorking, Surrey RH5 6NT, UK\label{aff145}
\and
Department of Astrophysical Sciences, Peyton Hall, Princeton University, Princeton, NJ 08544, USA\label{aff146}
\and
Cosmic Dawn Center (DAWN)\label{aff147}
\and
Niels Bohr Institute, University of Copenhagen, Jagtvej 128, 2200 Copenhagen, Denmark\label{aff148}
\and
Center for Cosmology and Particle Physics, Department of Physics, New York University, New York, NY 10003, USA\label{aff149}
\and
Center for Computational Astrophysics, Flatiron Institute, 162 5th Avenue, 10010, New York, NY, USA\label{aff150}}      

%
%
\abstract{
The \Euclid mission, designed to map the geometry of the dark Universe, presents an unprecedented opportunity for advancing our understanding of the cosmos through its photometric galaxy cluster survey. Central to this endeavor is the accurate calibration of the mass- and redshift-dependent halo bias (HB), which is the focus of this paper. We enhance the precision of HB predictions, which is crucial for deriving cosmological constraints from the clustering of galaxy clusters. Our study is based on the peak-background split (PBS) model linked to the halo mass function (HMF), and extends it with a parametric correction to precisely align with results from an extended set of $N$-body simulations carried out with the \og\ code. Employing simulations with fixed and paired initial conditions, we meticulously analyze the matter-halo cross-spectrum and model its covariance using a large number of mock catalogs generated with Lagrangian Perturbation Theory simulations with the \pinocchio\ code. This ensures a comprehensive understanding of the uncertainties in our HB calibration. Our findings indicate that the calibrated HB model is remarkably resilient against changes in cosmological parameters including those involving massive neutrinos. The robustness and adaptability of our calibrated HB model provide an important contribution to the cosmological exploitation of the cluster surveys to be provided by the \Euclid mission. This study highlights the necessity of continuously refining the calibration of cosmological tools like the HB to match the advancing quality of observational data. As we project the impact of our calibrated model on cosmological constraints, we find that, given the sensitivity of the \Euclid survey, a miscalibration of the HB could introduce biases in cluster cosmology analyses. Our work fills this critical gap, ensuring the HB calibration matches the expected precision of the \Euclid survey. 
}

%
%
%
%
   \titlerunning{\Euclid preparation. L. Calibration of the linear halo bias in $\Lambda(\nu)$CDM cosmologies}
   \authorrunning{Euclid Collaboration: Castro et al.}
   
   \maketitle
%
%
%
%
   
\section{\label{sc:Intro}Introduction}
The structure formation process in the Universe is hierarchical, with smaller structures collapsing and merging to form larger ones. Galaxy clusters, the most massive virialized objects in the Universe, lie at the apex of this hierarchy. They serve as valuable cosmological probes, offering insights into the growth of density perturbations and the geometry of the Universe~\citep[see, for instance,][for reviews]{Allen:2011zs,KravtsovBorgani:2012}.

The cosmological exploitation of cluster surveys is primarily based on number count analysis. This method involves comparing the observed number of clusters in a survey, as a function of redshift and of a given observable quantity, to the theoretical prediction of the halo mass function (HMF) within a cosmological model, thus enabling the derivation of constraints on cosmological parameters. Numerous studies have been conducted in this area~\citep[e.g.][]{Borgani:2001,Holder:2001db,DSDD:2009php,Hasselfield:2013wf,Planck:2013lkt,Bocquet:2014lmj,Mantz:2014paa,Planck:2015lwi,SPT:2018njh,DES:2020ahh,DES:2020cbm,Lesci:2020qpk}.

Complementing cluster count analysis is the cluster clustering statistics, which examines their spatial distribution in the Universe~\citep{Mana:2013qba,Castro:2016jmw,Baxter:2016jdq,DES:2020mlx,Lesci:2022owx,Sunayama:2023hfm,Romanello:2023obk,Fumagalli:2023yym}. The halo bias (HB) is a fundamental concept in this analysis since it reflects the ratio between the number overdensity of cluster and of the matter distribution. This relationship is expected to bring cosmological information through the mass- and redshift-dependence of the HB and to be linear on large scales, i.e., $\gtrsim 100 \, {\rm Mpc}$, to guarantee the scale-independence of the HB.

The \Euclid mission~\citep{EUCLID:2011zbd,Scaramella:2021,euclidoverview} is projected to provide significant advancements to cluster cosmology. \citet{Sartoris:2016} have forecasted that the combined cluster count and clustering analysis by the \Euclid mission will provide constraints on the amplitude of the matter power spectrum and the mass density parameters independent and competitive with other cosmological probes, underlining the potential of galaxy clusters as cosmological probes for ongoing and future missions.

At the heart of cluster cosmology are the theoretical models for the HMF and HB. Simplified models based on linear perturbation theory and spherical collapse have provided invaluable insights into the potential of cluster counts and clustering as cosmological probes~\citep[see, e.g.,][]{Press:1973iz,Bond:1990iw}. However, given the complexity and strongly non-linear nature of cluster formation dynamics, a refinement of these models to the precision level required by available and forthcoming surveys has to rely on cosmological simulations as the primary method to capture such complexity. Several studies have been dedicated to calibrating semi-analytical models for the HMF and HB, aiming to align these models' predictions with the results from extensive sets of simulations~\citep[see, for instance,][]{Sheth:1999mn, Sheth:1999su,Jenkins:2000bv,Warren:2005ey,Tinker:2008ff, Tinker:2010my,Bhattacharya:2010wy,Watson:2012mt,Despali:2015yla,Comparat:2017ejl, Euclid:2022dbc}. These simulations not only accurately describe the gravitational interactions that predominantly drive structure formation but also attempt to account for the effects of baryonic matter.

The influence of baryons, albeit a minor component in the Universe's overall composition, plays a significant role in the formation of structures, particularly in the context of these simulations~\citep{Cui:2014aga,Velliscig:2014,Bocquet:2015pva,Castro:2020yes}. Given the sensitivity of baryon evolution to the inclusion and modelling of astrophysical processes occurring at scales much smaller than those resolved in simulations, the modeling of baryonic feedback in hydrodynamical simulations remains a subject of active debate. At the scale of galaxy clusters, for instance, baryonic feedback is known to reorganize the mass density profile of halos without disrupting the structures, thereby altering the mass enclosed within a given radius compared to predictions from collisionless \textit{N}-body simulations. Owing to the substantially greater computational demands of hydrodynamical simulations, it has become standard practice to derive the HMF from gravitational \textit{N}-body simulations, with subsequent post-processing to account for baryonic effects~\citep[see, e.g.][]{Schneider:2015wta, Arico:2020lhq}. In this paper, we concentrate on the initial step of calibrating the HB using collisionless simulations. This approach is intended as a foundational phase, with the baryonic effects being integrated later, employing the methodology akin to that used for the HMF~\citep[see,][]{Castro:2020yes,Euclid:2023jih}. This strategy underscores our commitment to systematically exploring the cosmological parameter space, acknowledging the importance of baryonic effects while methodically building toward their inclusion in our analysis.

Systematic errors in the calibration of the HMF and HB can significantly impact the final cosmological constraints. Studies like those by~\citet{Salvati:2020exw}, \citet{Artis:2021tjj}, and \citet{Euclid:2022dbc} have highlighted how inaccuracies in theoretical models can propagate biases into cosmological parameter inferences. In response to these challenges, \citet{Euclid:2022dbc} presented a new, rigorously studied calibration of the HMF based on a suitably designed set of $N$-body simulations, offering the required accuracy to analyze \Euclid cluster count data.

Semi-analytical modeling typically starts with a simplified physical model, such as the peak-background split~\citep[PBS,][]{Mo:1995cs}, which is then extended and refined by adding more degrees of freedom. These additional degrees of freedom are subsequently fitted to simulations. Conceptually, PBS links the HB to the HMF by decomposing the density field into high- and low-frequency modes. The high-frequency modes that cross the collapsing barrier describe the collapse of structures. In contrast, the low-frequency modes modulate the density field fluctuations, thereby enhancing the number of peaks that cross the collapse threshold, therefore linking the clustering of collapsed objects with the local density field. Despite its qualitative consistency with simulations, PBS must enhance its quantitative precision, especially in the context of the \Euclid mission's requirements.

In this paper, we address the challenge of enhancing the accuracy of HB predictions to the level required to fully exploit the cosmological potential of the 2-point clustering statistics from the \Euclid photometric cluster survey. Our approach involves calibrating a semi-analytical model to quantify the discrepancies between PBS predictions and simulation results. This calibration aims to refine HB predictions, improving the reliability of cosmological parameter estimation derived from cluster counts and clustering.

This paper is organized as follows: we revisit the theoretical aspects used in this paper in Sect.~\ref{sec:theory}. In Sect.~\ref{sec:methodology}, we describe the methodology used in our analysis. We present the HB model and its calibration in  Sect.~\ref{sec:results}, along with an assessment of our model's impact in a forecast \Euclid cluster cosmology analysis. Final remarks are made in Sect.~\ref{sec:conclusions}. The implementation of our model is publicly available in~\faGithub~\href{https://github.com/TiagoBsCastro/CCToolkit}{https://github.com/TiagoBsCastro/CCToolkit} and presented in Sect.~\ref{sec:dataavailability}.

\section{\label{sec:theory} Theory}

\subsection{The halo mass function}

The differential HMF is given by
\begin{equation}
	\frac{\de n}{\de M} \de M = \frac{\rho_{\rm m}}{M} \, \nu f(\nu) \, \de \ln \nu\,,
	\label{eq:hmf}
\end{equation}
where $n$ is the comoving number density of halos with mass in the range $[M, M+\de M]$, $\nu$ is the peak height, the function $\nu f(\nu)$ is known as the multiplicity function, and $\rho_{\rm m}$ is the comoving cosmic mean matter density
\begin{equation}
    \rho_{\rm m} = \frac{3\,H_0^2\,\Omega_{\rm m, 0}}{8\pi G}\,,
\end{equation}
where $H_0$ and $\Omega_{\rm m, 0}$ are the current value of the Hubble parameter and the matter density parameter, and $G$ is the gravitational constant. The peak height is defined as $\nu=\delta_{\rm c}/\sigma(M, z)$, where $\delta_{\rm c}$ is the critical density for spherical collapse~\citep{peebles2020large} and $\sigma^2(M, z)$ is the filtered mass variance at redshift $z$, so that it measures how rare a halo is. The mass variance is expressed in terms of the linear matter power spectrum $P_{\rm m}(k,z)$ as
\begin{equation}
	\sigma^2(M,z) = \frac{1}{2\pi^2} \int_0^\infty \de k\,k^2\,P_{\rm m}(k,z)\,W^2 \left(k\,R \right)\,,
\label{eq:massvariance}
\end{equation}
where $R(M)= \left[3\,M\,/\,(4\,\pi\,\rho_{\rm m})\right]^{1/3} $ is the Lagrangian radius of a sphere containing the mass $M$, and $W(k\,R)$ is the Fourier transform of a  top-hat filter of radius $R$.

The multiplicity function is considered universal if its cosmological dependence is solely through the peak height. However, numerous studies based on $N$-body simulations have challenged this assumption. These analyses reveal that, while the initial approximation of HMF universality is generally valid, systematic deviations from this universality become evident at late times in the universe's evolution. This deviation has been demonstrated in various independent investigations, each indicating a nuanced understanding of the HMF's behavior~\citep[e.g.,][]{Crocce:2009mg, Courtin:2010gx, Watson:2012mt, Diemer:2020rgd, Ondaro-Mallea:2021yfv,Euclid:2022dbc}.

The non-universality of the HMF  is affected by both the halo definition and the residual dependence of $\delta_{\rm c}$ on cosmology. Various studies have shown this dependence, including~\citet{Watson:2012mt}, \citet{Despali:2015yla}, \citet{Diemer:2020rgd}, and \citet{Ondaro-Mallea:2021yfv} for the dependence on the halo definition, and~\citet{Courtin:2010gx} for the cosmology dependence of $\delta_{\rm c}$. In our study, we define halos as spherical overdensities (SO) with an average enclosed mean density equal to $\Delta_{\rm vir}(z)$ times the background density, where $\Delta_{\rm vir}(z)$ is the non-linear density contrast of virialized structures as predicted by spherical collapse~\citep{Eke:1996ds,Bryan_1998}. The multiplicity function for halo masses computed at the virial radius has been shown to preserve universality better than other commonly assumed definition of halo radii ~\citep{Despali:2015yla,Diemer:2020rgd,Ondaro-Mallea:2021yfv}. As for $\delta_{\rm c}$, we use the fitting formula introduced by~\citet{Kitayama:1996ne} that ignores the effect of massive neutrinos; however, for the adopted values for total neutrino masses in this work, the fitting formula is still percent level accurate~\citep{LoVerde:2014rxa}.

In this paper, we use the HMF presented in~\citet{Euclid:2022dbc}
\begin{equation}
	\label{eq:mult}
	\nu\,f(\nu)=A(p,q) \sqrt{\frac{2a\nu^2}{\pi}} \mathrm{e}^{-a\nu^2/2} \left(1+ \frac{1}{(a\nu^2)^p} \right) (\nu\sqrt{a})^{q-1} \,,
\end{equation}
where the parameters $\{a, p, q\}$ depend on background evolution and power spectrum shape as
\begin{align}
	\label{eq:parevol1}
	a &= a_R \,  \Omega_{\rm m}^{a_z}(z) \,,\\
	p &= p_1 + p_2 \,  \left( \frac{\de \ln \sigma}{\de \ln R} + 0.5 \right) \,,\\
	q &= q_R \, \Omega_{\rm m}^{q_z}(z) \,,
\end{align}
where $\Omega_{\rm m}$ is the fractional density of matter in the Universe as a function of redshift, encompassing both baryonic and dark matter contributions
\begin{align}
a_R &= a_1 + a_2 \,  \left( \frac{\de \ln \sigma}{\de \ln R} + 0.6125 \right)^2 \,,\\
q_R &= q_1 + q_2 \,  \left( \frac{\de \ln \sigma}{\de \ln R} + 0.5 \right) \,.
\label{eq:parevol2}
\end{align}

Lastly, the normalization parameter $A$ is not a free parameter but a function of the other parameters
\begin{equation}
	A(p,q) = \Bigg\{ \frac{2^{-1/2-p+q/2}}{\sqrt{\pi}} \, \left[ 2^p \, \Gamma\left(\frac{q}{2}\right)+ \Gamma\left(-p+\frac{q}{2}\right)  \right] \Bigg\}^{-1}\, ,
\label{eq:Anorm}
\end{equation}
where $\Gamma$ denotes the Gamma function. The adopted values for the HMF parameters are presented in Table 4 of~\citet{Euclid:2022dbc} and depend on the halo-finder used. In this work, we mostly use the \rockstar\ calibration. The \subfind\ calibration is also used in Sect.~\ref{sec:hf} to assess the impact of the halo-finder in our model.

\subsection{The linear halo bias}

The overdensity of halos of mass $M$ at the position $\textbf{r}$ at redshift $z$
\begin{equation}
\delta_{\rm h}(\textbf{r},M,z)=n(\textbf{r},M,z)/\bar{n}(M,z)-1\,,
\end{equation}
is expressed in terms of the corresponding local number halo number density, $n(\textbf{r},M,z)$, and of the cosmic mean number density of such halos, $\bar{n}(M,z)$. In linear theory, it 
is related to the matter density contrast $\delta_{\rm m}(\textbf{r},z)$ as
\begin{equation}
    \delta_{\rm h}(\textbf{r},M,z)=b(M,z)\,\delta_{\rm m}(\textbf{r},z)+\epsilon(\textbf{r},M,z)\,,
    \label{eq:bias}
\end{equation}
where $b(M,z)$ is the linear halo bias and $\epsilon$ is a stochastic term that in the following we assume to be associated with shot-noise. 

It follows from Eq.~\eqref{eq:bias} that the halo-halo, $P_{\rm h}$, and halo-matter power spectrum, $P_{\rm hm}$, are written as a function of the linear matter power spectrum, $P_{\rm m}$, for sufficiently large scales, as
\begin{align}
P_{\rm h}(k, M,z)&=b^2(M,z)\,P_{\rm m}(k, z)+P_{\rm SN}\,,\\
P_{\rm hm}(k, M,z)&=b(M,z)\,P_{\rm m}(k, z)\,,
\end{align}
where $P_{\rm SN}$ represents the shot-noise component. Under the assumption that halos offer a discrete Poisson sample of the underlying continuous matter density field, $P_{\rm SN}$ denotes a shot-noise component commonly assumed to be equivalent to the Poisson term, $P_{\rm SN}=1/\bar{n}$, where $\bar{n}$ is the mean number density of tracers. On the other hand, halos are known to correspond with high-density peaks of the underlying matter distribution. Therefore, they are expected not to provide a purely Poissonian sampling of this continuous density field. In fact, \citet{CasasMiranda:2001ym} and \citet{Hamaus:2010im} showed that positive and negative corrections to Poisson shot-noise are expected for low- and high-mass halos. In this paper, we parameterize $P_{\rm SN}$ as
\begin{equation}
    P_{\rm SN}=\frac{1-\alpha}{\bar{n}}\,,
    \label{eq:sn}
\end{equation}
where $\alpha$, a fitting parameter that we calibrate through simulations, controls the deviation from the assumption of Poisson noise.

Assuming the universality of the HMF,~\citet{Mo:1995cs} derived the HB $b(M, z)$ directly from the HMF by following the PBS framework. The PBS prediction for the HB as a function of the peak height $\nu$ reads
\begin{equation}
b_{\rm PBS}(\nu) = 1 - \frac{1}{\delta_{\rm c}} \frac{\de \ln{\nu\,f(\nu)}}{\de \ln{\nu}}\,.
\label{eq:biasps}
\end{equation}

Although the PBS provides an estimate of the bias that presents the correct qualitative behavior,~\citet{Tinker:2010my} claims a relatively poor performance of the PBS in reproducing results from $N$-body simulations, with an accuracy of about 10--20 percent. 

Given the correct qualitative behavior of the PBS prescription, in this paper, we aim to improve the prediction of the bias by calibrating a model for the bias as a function of $b_{\rm PBS}$ -- i.e., we assume Eq.~\eqref{eq:biasps} to be valid, with the HMF~\eqref{eq:mult} -- and model its difference with respect to the simulations.

\section{\label{sec:methodology}Methodology}

\subsection{\label{sec:sims}Simulations}

\subsubsection{\label{sec:nbody}$N$-body simulations}

Table~\ref{tab:cosm} presents the adopted values for the matter density parameter and baryonic density parameter at redshift 0 ($\Omega_{\rm m, 0}$ and $\Omega_{\rm b, 0}$), the dimensionless Hubble parameter $h$, the spectral index of the primordial power spectrum $n_{\rm s}$, and the amplitude of matter density fluctuations on scales of $8\mpc$ $\sigma_8$ for the $N$-body simulations used in this work. We extended the set of \piccolo\ simulations introduced and used by~\citet{Euclid:2022dbc} to calibrate the HMF model. We maintain the same technical configurations as the original \piccolo\ simulations and refer to the above-mentioned HMF paper for further details while summarizing the main aspects. The set comprises $69$ cosmological boxes, each with a comoving size of $2000\,\mpc$, and $4\times 1280^3$ dark matter particles. The simulations were conducted using \og, with initial conditions generated by \monofonic~\citep{Michaux:2020yis}, based on third-order Lagrangian Perturbation Theory (3LPT) at a redshift of $z=24$. The adopted gravitational softening is equivalent to one-fortieth of the mean inter-particle distance.

The original \piccolo\ set of simulations included nine different choices for cosmological parameters, which were randomly chosen from the $95$ percent confidence level hyper-volume of the joint SPT and DES cluster abundance constraints~\citep{DES:2020cbm}. Those represent the cosmologies $C0$ to $C8$ in Table~\ref{tab:cosm}. To further stress our modeling and guarantee its robustness, we also add the cosmologies $C9$ and $C10$, which sample the $(\Omega_{\rm m,0},\sigma_8)$ plane in the direction orthogonal to the degeneracy direction of the constraints from~\citet{DES:2020cbm}, and in significant tension with such constraints. For each cosmology, two white-noise realizations were created to generate initial conditions. For each noise realization, a pair was generated by fixing the amplitudes of the Fourier modes of the density fluctuation field and pairing the phases~\citep{Angulo:2016hjd}, except for the reference $C0$ model, which had 20 realizations.

We further add three simulations with Einstein--de Sitter cosmology (EdS, i.e., $\Omega_{\rm m} = 1$ and $\Omega_{\rm \Lambda} = 0$) with power-law initial matter power spectrum  $P_{\rm m}(k) \propto k^{n_{\rm s}}$ with $n_{\rm s} \in \{-1.5, -2.0, -2.5\}$. Those simulations, which have the same box size and the same number of particles as the \piccolo\ set, are only instrumental for the modeling and will not be used for the calibration as they are far from the regime used to calibrate the model of~\citet{Euclid:2022dbc}.

Lastly, we carried out three pairs of simulations with massive neutrinos, again using the same box size as the \piccolo\ set. Each pair has a total neutrino mass $\sum m_\nu\in\{0.15, 0.30, 0.60\}\,{\rm eV}$. The simulation set-up for the neutrino simulations is the same used for the \og\ simulations extensively validated in~\citet{Euclid:2022qde}. The baseline cosmological parameters is the $C0$ set, with $\Omega_{\nu, 0}$ subtracted from $\Omega_{\rm m, 0}$. The simulations have the same primordial amplitude $A_{\rm s}$, resulting in a lower $\sigma_8$ for increasing neutrino mass (see Table \ref{tab:cosm}). 

The initial conditions (ICs) for the neutrino simulations were generated using the {\tt FastDF}~\citep{Elbers:2022xid} implementation to \monofonic~\citep{Michaux:2020yis}.\footnote{\url{https://bitbucket.org/ohahn/monofonic}} The forked repository with the {\tt FastDF} integration can be found here.\footnote{\url{https://github.com/wullm/monofonic}} We employed the same number of neutrino particles as the number of grid resolution elements used for the cold dark matter particles. The total neutrino mass specified is attributed to a single massive neutrino species.

\begin{table}
	\centering
	\caption{The cosmological parameters of the \piccolo\ set of simulations.}
	\label{tab:cosm}
	\begin{tabular}{c|c|c|c|c|c}
		\hline
		Name & $\Omega_{\rm m,0}$ & $h$ & $\Omega_{\rm b,0}$ & $n_{\rm s}$ & $\sigma_8$ \\\hline
		$C0$ & $0.3158$ & $0.6732$ & $0.0494$ & $\phantom{-}0.9661$ & $0.8102$ \\
		$C1$ & $0.1986$ & $0.7267$ & $0.0389$ & $\phantom{-}0.9775$ & $0.8590$ \\
		$C2$ & $0.1665$ & $0.7066$ & $0.0417$ & $\phantom{-}0.9461$ & $0.8341$ \\
		$C3$ & $0.3750$ & $0.6177$ & $0.0625$ & $\phantom{-}0.9778$ & $0.7136$ \\
		$C4$ & $0.3673$ & $0.6353$ & $0.0519$ & $\phantom{-}0.9998$ & $0.7121$ \\
		$C5$ & $0.1908$ & $0.6507$ & $0.0527$ & $\phantom{-}0.9908$ & $0.8971$ \\
		$C6$ & $0.2401$ & $0.8087$ & $0.0357$ & $\phantom{-}0.9475$ & $0.8036$ \\
		$C7$ & $0.3020$ & $0.5514$ & $0.0674$ & $\phantom{-}0.9545$ & $0.8163$ \\
		$C8$ & $0.4093$ & $0.7080$ & $0.0446$ & $\phantom{-}0.9791$ & $0.7253$ \\
        \hline
        $C9$  & $0.1530$ & $0.6660$ & $0.0408$ & $\phantom{-}0.9661$ & $0.6140$ \\
        $C10$ & $0.4280$ & $0.7300$ & $0.0492$ & $\phantom{-}0.9661$ & $0.9000$ \\
        \hline
        ${\rm EdS}0$ & $1.0000$ & $0.7000$ & $0.0000$ & $-1.5000$ & $0.5000$ \\
        ${\rm EdS}1$ & $1.0000$ & $0.7000$ & $0.0000$ & $-2.0000$ & $0.5000$ \\
        ${\rm EdS}2$ & $1.0000$ & $0.7000$ & $0.0000$ & $-2.5000$ & $0.5000$ \\
        \hline 
        $\nu$-$150$ & $0.3158$ & $0.6732$ & $0.0494$ & $\phantom{-}0.9661$ & $0.8036$ \\
        $\nu$-$300$ & $0.3158$ & $0.6732$ & $0.0494$ & $\phantom{-}0.9661$ & $0.7709$ \\
        $\nu$-$600$ & $0.3158$ & $0.6732$ & $0.0494$ & $\phantom{-}0.9661$ & $0.7123$ \\
        \hline
		\hline
	\end{tabular}
 \tablefoot{The parameters of the $C0$ to $C8$ models have been uniformly drawn from the $95$ per cent confidence level hyper-volume of the cluster abundance constraints presented in~\citet{DES:2020cbm}. The parameters for the $C9$ and $C10$ were specifically selected to sample the parameter space in the direction orthogonal to the direction of degeneracy in the $(\Omega_{{\rm m},0},\sigma_8)$ plane of the constraints of~\citet{DES:2020cbm}, and in significant tension with them.}
\end{table}

\subsubsection{\label{sec:pin}Approximate methods: \pinocchio}

In this paper, we also analysed $200$ ($100$ pairs with each pair having fixed amplitudes and paired phases) halo catalogs simulated with the approximate LPT-based method implemented in the \pinocchio\ code~\citep{Monaco:2001jg,Monaco:2013qta,Munari:2016aut}. All these simulations have been carried out under the assumption of the $C0$ cosmological parameters. The rationale for this extra set of simulated catalogs is to model the impact of fixing and pairing the Fourier mode amplitudes in the ICs on the cluster clustering.

\subsection{\label{sec:halocats}Halo finders}

\citet{Euclid:2022dbc} showed that the halo-finder adopted for the analysis of the $N$-body simulations can significantly alter the HMF. To understand if the halo-finder also impacts the HB, we selected two algorithms to extract halo catalogs: \rockstar~\citep{Behroozi:2011ju}\footnote{\url{https://bitbucket.org/gfcstanford/rockstar}} and \subfind~\citep{Springel:2000qu,Dolag:2008ar,Springel:2020plp}. Although all these algorithms rely on the SO method to define halo boundaries, they differ in the method used to identify the center from which the spheres are grown and the criteria to classify between structures and sub-structures. 

\rockstar\ divides the simulation volume into 3D friend-of-friends (FOF;~\citealp[see, for instance,][]{Davis:1985rj}) groups and runs a recursive 6D FOF algorithm on each group to create a hierarchy of FOF subgroups. Halo centers are determined by averaging the positions of particles in the innermost subgroup. To improve consistency, we apply the \consistent\ algorithm, which dynamically tracks halo progenitors, to the extracted \rockstar\ catalogs as demonstrated in~\citet{Behroozi:2011js}. \subfind\ also determines halo centers using a parallel implementation of the 3D FOF algorithm but directly assigns it to the particle with the lowest gravitational potential. 

Among the halo-finders studied by~\citet{Euclid:2022dbc}, \rockstar\ and \subfind\ are good representatives of the heterogeneity of possible results as they are close to the extremes, with \subfind\ suppressing the abundance of objects more massive than $10^{13}\,\msun$ by roughly 10 percent. See~\citet{Knebe:2011rx} for a more detailed comparison between the halo-finding algorithms.

\subsection{\label{sec:measuring}Measuring the halo bias}

To measure the HB, we bin the halo distribution in $\log_{10}( M_{\rm vir}/\msun)$ with equispaced width intervals of $0.1$ dex at each redshift. If the number of halos inside a bin was less than $10\,000$, we merged it with its neighbor to avoid having bins where the power-spectrum measurements were primarily dominated by shot noise. We measured the cross-spectrum $P_{\rm hm}$ between the halos in each mass bin and the matter distribution traced by particles. The bias is then obtained as the ratio between this cross-spectrum and the matter power spectrum $P_{\rm m}(k)$. The matter density field was computed at the initial conditions and rescaled according to the linear growth factor for the simulations without massive neutrinos. For the simulations with massive neutrinos, the matter density field was built from the particle data from the same snapshot from which the halo catalog was extracted to consider the scale-dependent growth factor. We used the \pylians\footnote{\url{https://github.com/franciscovillaescusa/Pylians3}} Python libraries to construct the density field and compute power spectra on a $1024^3$ piecewise cubic spline mesh grid. \pylians\ averages the power spectra in $k$-space shells with the width given by the fundamental mode of the box, $k_{\rm f}\equiv 2\pi/L$. Following~\citet{Castro:2020yes}, we only considered modes with $k$ values smaller than $0.05\,\mpcinv$ to measure the HB, to ensure the validity of the linear approximation. The maximum $k$ used for the measurements corresponds to the $16^{\rm th}$ harmonic of the box and is much smaller than the Nyquist frequency of the grid used to compute the power spectrum.

To calculate the bias for each mass bin, we used the ratio of the halo-matter cross-spectrum and the matter power spectrum
\begin{equation}
b_{i,j}^{\rm sim}=P_{\rm hm}(k_j)_i/P_{\rm m}(k_j)\,,    
\label{eq:bsim}
\end{equation}
where $i$ and $j$ are the mass bin and the $k$-shell indexes, respectively.

\subsection{\label{sec:likelihood}Halo bias calibration}

We used a Bayesian approach with uninformative uniform priors on all parameters to fit our model for the linear HB parameters to simulation results. The best fits were obtained using the Dual Annealing method to find the posterior maximum as implemented by \citet{Virtanen_2020}, and covariance between parameters was estimated using \pymc~\citep{salvatier2016probabilistic}.\footnote{\url{https://www.pymc.io}} The No-U-Turn sampler (NUTS)~\citep{hoffman2014no} was automatically assigned internally by \pymc\ to sample the likelihood.

We assume a Gaussian likelihood for the bias, since the power spectrum estimation for a Gaussian field realization follows a $\chi^2$-distribution when averaged over a shell. Since the number of modes $N_{k}$ inside the shell increases rapidly with $k$, this distribution approaches a Gaussian distribution by the central limit theorem. However, the number of modes is small for the first bins, and deviation from the Gaussian approximation is evident. To avoid this issue, we re-bin the measurement of the first 3 $k$-bins by merging them, ensuring that the bin with the fewest modes still contains 117 modes to recover the Gaussian approximation's validity.

Note that differently than~\citet{Castro:2020yes}, we used ICs with fixed amplitudes. Therefore, the distribution of the simulated bias is not approximately a ratio of two Gaussian distributions but approximately Gaussian itself since the denominator of Eq.~\eqref{eq:bsim} is not a random variable. The variance of the shell-average halo-matter cross-spectrum on simulations with random Gaussian initial conditions is given by
\begin{equation}
\left(\frac{\sigma_{P_{\rm hm}}}{P_{ \rm m}}\right)^2 = \frac{1}{N_k} \,\left( b^2 + \frac{1-\alpha}{\bar{n}\,P_{\rm m}} \right)\,, 
\label{eq:cross_var}
\end{equation}
where we assume that the shot-noise contribution to the halo power spectrum follows Eq.~\eqref{eq:sn} and a linear HB $b$. However, \citet{Zhang:2021hrq} showed that the predictions for random Gaussian initial conditions overestimate the variance observed for biased tracers on simulations with fixed amplitudes. Therefore, we modify Eq.~\eqref{eq:cross_var} as follows
\begin{equation}
\left(\frac{\sigma_{P_{\rm hm}}}{P_{ \rm m}}\right)^2 \equiv \sigma_{b}^2 = \frac{1}{N_k} \,\left( \beta\,b^2 + \frac{1-\alpha}{\bar{n}\,P_{\rm m}} \right) + b^2\,\sigma_{\rm sys}^2 \,.
\label{eq:cross_var_2}
\end{equation}
Here, $\beta$ and $\sigma_{\rm sys}$ are parameters we marginalize over, which control the variance suppression and the relative error due to the limited accuracy of the HMF used as the backbone for the PBS prescription. We note that, based on the halo model~\citep[see, for instance][]{Cooray:2002dia}, one should expect a value for $\beta$ close to zero as in the limit where all halos are considered, the shot-noise term on the right-hand side of Eq.~\eqref{eq:cross_var_2} tends to zero, so that one should recover the matter power spectrum that by construction has zero variance. Furthermore, the HMF presented in~\citet{Euclid:2022dbc} was shown to have percent-level accuracy; thus, $\sigma_{\rm sys}$ is expected to assume similar values during the calibration, presuming the PBS framework is valid. However, it is crucial to note that should $\sigma_{\rm sys}$ significantly deviate from zero, such an occurrence could indicate a potential violation or limitation within the PBS framework, underscoring the necessity for careful interpretation of these parameters.

We obtain the total log-likelihood by summing up all mass bins, modes, redshifts, and simulations. Following~\citet{Euclid:2022dbc}, we use the redshifts $z \in \{2.00, 1.25, 0.90, 0.52, 0.29, 0.14, 0.0\}$, translating to a time-spacing of about $1.7$ Gyr. This spacing is larger than the characteristic dynamical time of galaxy clusters and effectively suppresses the correlation between the results of different snapshots. Similarly, we assume that the correlations between different mass bins and modes are negligible. This assumption is justified by linear theory, which posits that different modes evolve independently in the linear regime, thus minimizing their mutual influence. In our analysis, we have three fitting parameters: $\alpha$, $\beta$, and $\sigma_{\rm sys}$, in addition to the parameters of the bias model that are subject to calibration (to be discussed in Sect.~\ref{sec:model}). This approach allows for a comprehensive calibration of the bias model, taking into account the shot-noise correction $\alpha$, the suppression of variance $\beta$, and the systematic uncertainties $\sigma_{\rm sys}$ inherent in our method.

\subsection{\label{sec:forecast}Forecasting \Euclid's cluster counts and cluster clustering observations}

To understand the impact of the HB calibration on cosmological constraints, it is important to realistically forecast the cosmological information to be extracted from the \Euclid photometric cluster survey. For this purpose, we first quantify the impact of the HB on cluster counts and cluster clustering analyses. More precisely, the HB enters the modeling of the cluster counts covariance, which we compute analytically following the model of \citet{Hu:2002we}, as validated in \citet{Euclid:2021api}. Regarding the cluster clustering, the HB enters both in the computation of the mean value (power spectrum or two-point correlation function), and in the associated covariance matrix; in this work, we test the effect on the real-space two-point correlation function and its covariance, following the model presented and validated in \citet{Euclid:2022txd}. 

After assessing the impact on the two statistics, we forecast how the accuracy of the HB calibration propagates on the cosmological constraints obtained by cluster counts and cluster clustering experiments. We generate synthetic cluster abundance data as described in Section 2.5 of \citet{Euclid:2022dbc}, assuming the HMF calibrated and the HB of this work as benchmarks. Through a likelihood analyses, we constrain the cosmological parameters $\Omega_{\rm m,0}$ and $\sigma_8$, and the mass-observable relation (MOR) parameters $A_\lambda, B_\lambda, C_\lambda, D_\lambda$ \citep[see Section 2.5 of][]{Euclid:2023jih}, assuming flat priors for all the parameters. The MOR parameters describe the optical richness $\lambda$ distribution as a function of the halo mass according to
\begin{align}
	\langle \ln \lambda | M_{\rm vir}, z  \rangle =& \ln A_\lambda + B_\lambda \ln{\left(\frac{M_{\rm vir}}{3\times10^{14}\,h^{-1} M_\odot}\right)} \nonumber\\ 
	& + C_\lambda \ln{\left(\frac{H(z)}{H(z=0.6)}\right)}\,,
	\label{eq:lambda}
\end{align}
where $H(z)$ denotes the Hubble parameter at redshift $z$. The range for richness $\lambda$ is considered to be between 20 and 2000, with the variation in logarithmic richness for a given virial mass and redshift expressed as a log-normal scatter
\begin{equation}
\sigma_{\ln \lambda | M_{\rm vir}, z}^2 = D_\lambda^2\, .
\end{equation}
The reference values for the parameters are $A_\lambda = 37.8$, $B_\lambda = 1.16$, $C_\lambda = 0.91$, and $D_\lambda = 0.15$. These values have been obtained refitting the parameters presented in~\citet{DES:2015mqu} to the virial mass definition, with the assumption that halo profiles follow an NFW profile with a mass-concentration relationship as outlined in~\citet{Diemer:2018vmz}. The parameter values adopted in this work are consistent with the model presented by~\citet{Castignani:2016lvp} to assign cluster membership probabilities to galaxies from photometric surveys.

We perform the analysis on the synthetic catalogs, comparing them with the predictions made by using our bias and the one of \citet{Tinker:2010my}, and compare the resulting posteriors with two estimators: we quantify the broadening/tightening of the posterior's amplitude by computing the difference of the figure of merit~\citep[$\Delta {\rm FOM}$; see][]{Huterer:2000mj,Albrecht:2006um} in the $\Omega_{\rm m,0}$\,--\,$\sigma_8$ plane, and the shift of the posterior's position by computing the posterior agreement \citep{SPT:2018njh}, which determines whether the difference between two posterior distributions is consistent with zero.

\section{\label{sec:results}Results}

\subsection{\label{sec:hmf}Calibration of the halo bias}

\subsubsection{\label{sec:fixedandpaircov}Biased tracers statistics on fixed and paired simulations}

Before modeling and calibrating the HB to the simulations, we investigate the impact of the variance suppression technique on the halo matter cross-spectrum. 

\begin{figure}
    \centering
    \includegraphics[width=\columnwidth]{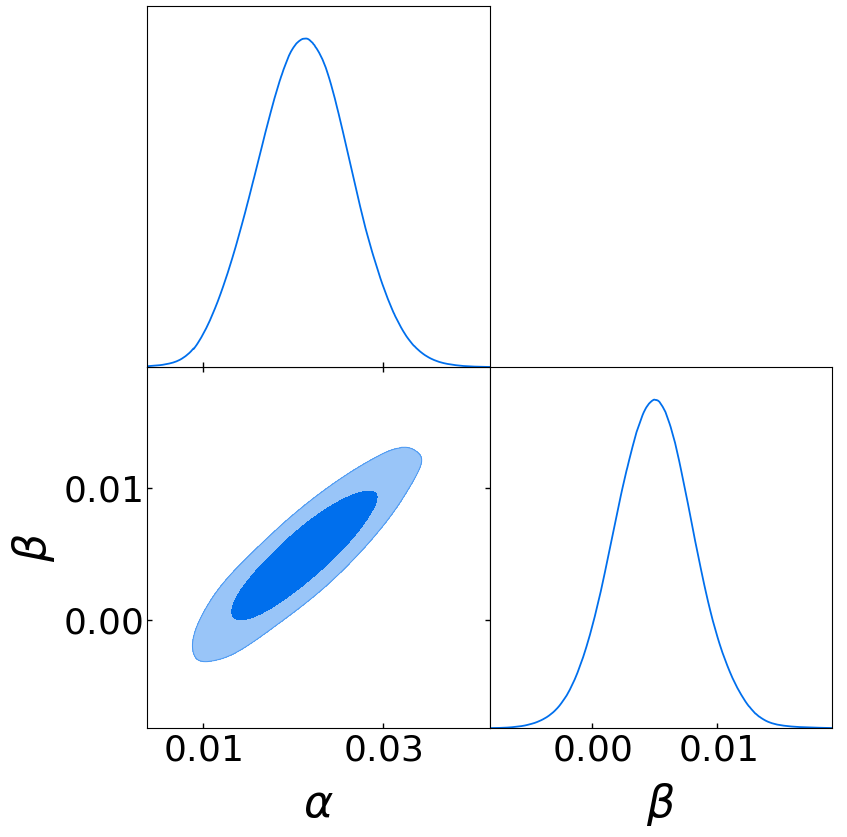}
    \caption{The constraints on the parameters in Eq.~\eqref{eq:cross_var_2} fitted to the unbiased standard deviation of the $200$ \pinocchio\ mocks with $C0$ cosmology. We fitted $\alpha$ and $\beta$ considering $k\leq0.05\,\mpcinv$ assuming a Gaussian likelihood with error bar estimated from the measurements for $0.05\leq k/\mpcinv\leq 0.2$ assuming it to be constant in $k$ and equivalent to the unbiased standard deviation.}
    \label{fig:f&pchain}
\end{figure}

In Fig.\ref{fig:f&pchain}, we present the constraints on the parameters in Eq.~\eqref{eq:cross_var_2} fitted to the unbiased standard deviation of the $200$ \pinocchio\ mocks with $C0$ cosmology. We fixed $\sigma_{\rm sys}$ to zero, as this exercise does not involve modeling errors on the bias. The parameters $\alpha$ and $\beta$ were fitted using only modes with $k\leq0.05\,\mpcinv$, under the assumption of a Gaussian likelihood. We estimated the error bars from the measurements within the range $0.05\leq k/\mpcinv\leq 0.2$, treating it as constant and equivalent to the unbiased standard deviation. This approach prevents the estimation of the expectation of the mean and the error from using the same data. Figure~\ref{fig:f&pchain} reveals a positive correlation between the parameters, with both assuming positive values. As expected, the posterior for the $\beta$ parameter peaks close to zero, validating the effectiveness of the fixed amplitudes technique in reducing the variance of biased tracer statistics. Notably, the small value of $\alpha$ found in our analysis suggests that the shot noise is well modeled as Poissonian to within 1--3 percent. 

\begin{figure*}
    \centering
    \includegraphics[width=0.99\textwidth]{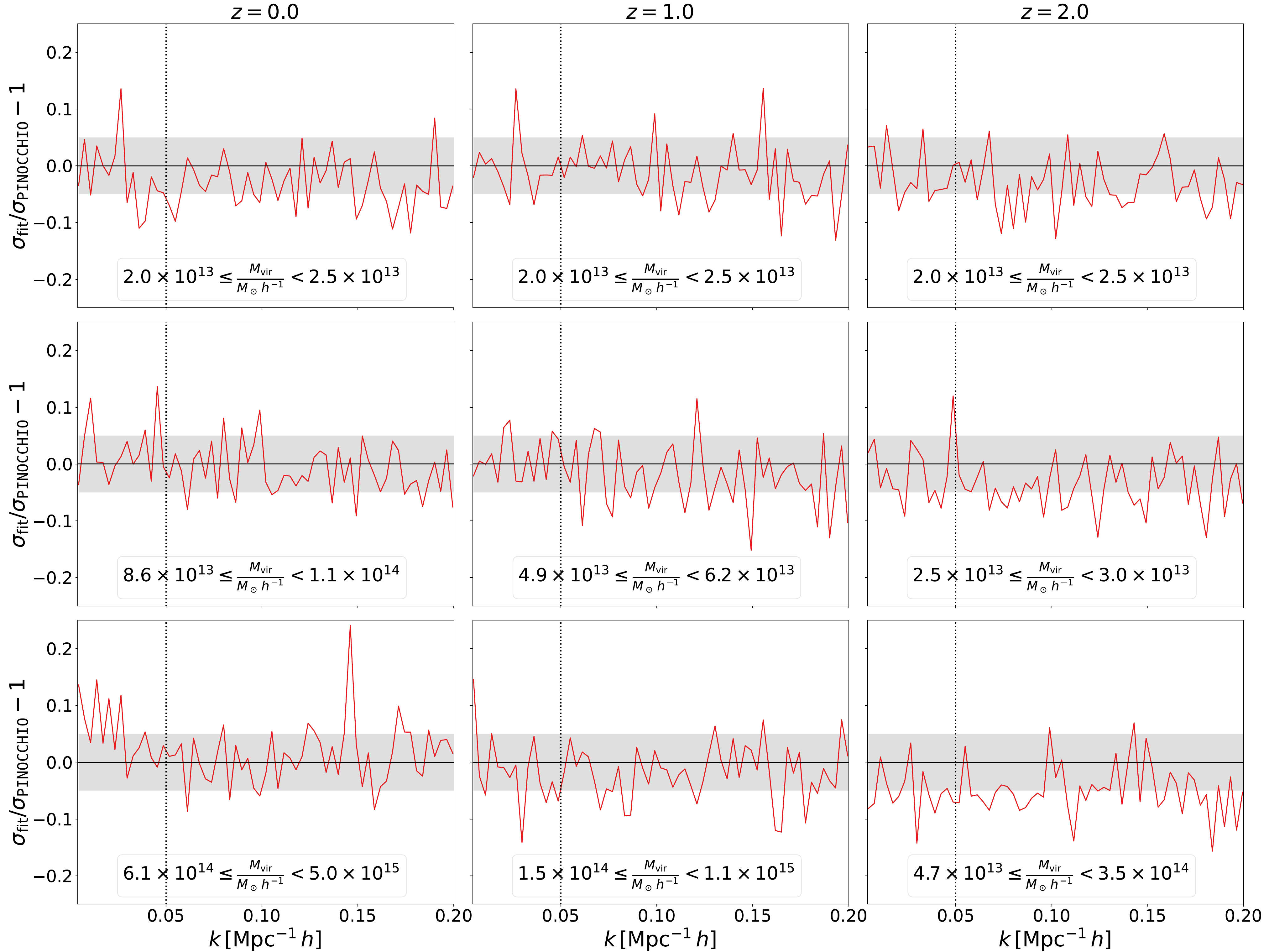}
    \caption{The relative difference between Eq.~\eqref{eq:cross_var_2} best-fit and the unbiased standard deviation of the \pinocchio\ measurements. Different columns are for different redshifts, and the corresponding mass bins are shown in each panel. The vertical dotted line demarcates two distinct sets of measurements: those to the left of the line were utilized as data points in the parameter fitting process, while the scatter of the points to the right was analyzed to estimate the variance. The grey regions highlight areas within a $5\%$ deviation from the expected values.}
    \label{fig:f&pvar}
\end{figure*}

In Fig.~\ref{fig:f&pvar}, we present the relative difference between Eq.~\eqref{eq:cross_var_2} computed with the best-fit values of $\alpha$ and $\beta$, and the unbiased standard deviation of the measurements, for different mass bins and redshift values. We present the results for three values $z \in {0.0, 1.0, 2.0}$ spanning the redshift range of interest, while for the masses, we present the first, the intermediate, and the last occupied bin for each redshift. We observe that the residuals of the fit always oscillate around zero, with no statistically significant mass, redshift, or $k$ dependence.

\begin{figure*}
    \centering
    \includegraphics[width=0.99\textwidth]{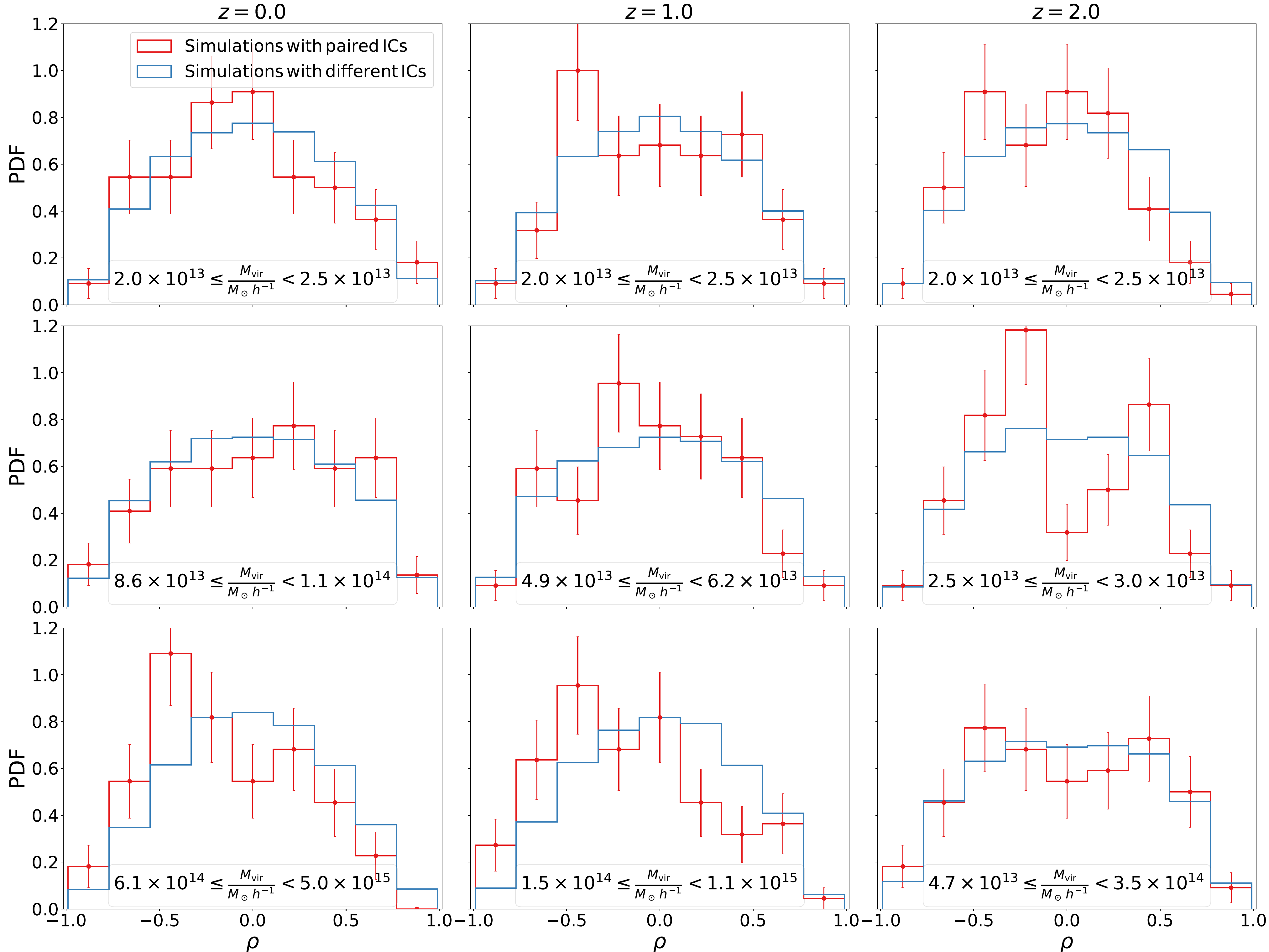}
    \caption{The correlation coefficient $\rho$ between the measurements of the halo-matter power spectrum, $P_{\rm hm}(k)$ in different simulations for $k \leq 0.05\,\mpcinv$, as a function of halo mass bin and redshift. The red histograms show the distribution of the correlation coefficient between a simulation and its paired realization, while blue histograms are for the correlation coefficient between simulations with uncorrelated white-noise realizations.}
    \label{fig:f&pcor}
\end{figure*}

In Fig.~\ref{fig:f&pcor}, we present the Pearson correlation coefficient $\rho$ between the measurements of the halo-matter cross-power spectrum on a given simulation and its paired realization. The correlation coefficient between two random variables $X$ and $Y$ is defined as
\begin{equation}
\rho(X, Y) = \frac{\left\langle (X - \langle X \rangle)(Y - \langle Y \rangle) \right\rangle}{\sigma_X\,\sigma_Y}\,,
\end{equation}
where $\sigma_X$ and $\sigma_Y$ are the standard deviation of the random variables $X$ and $Y$. The cross-power spectrum was measured for $k \leq 0.05\,\mpcinv$, for different ranges of halo masses (as reported in each panel) and redshift values (different columns).  We also present the correlation coefficient between simulations that assumes uncorrelated white-noise realizations for comparison. We note that paired simulations do not show a statistically significant difference in their correlation with respect to simulations that assume uncorrelated noise realizations. The same conclusion is obtained by running a p-value test on all mass and redshift bins. This result justifies the assumption that different simulations are, in fact, independent, even if they have the same amplitudes and paired phases. This conclusion aligns with the claims of~\citet{Villaescusa-Navarro:2018bpd} that variance suppression techniques are unlikely to affect the halo abundance distribution. As the shot-noise term in Eq.~\eqref{eq:cross_var_2} dominates over the other terms for our sample selection, one could anticipate the independence of the bias results from fixed and paired simulations due to the independence of abundance fluctuations and modes paring.  

Lastly, we use the \pinocchio\ catalogs to assess the impact of neglecting the correlation between different mass bins and Fourier modes on the calibration likelihood presented in Sect.~\ref{sec:likelihood}. We measured the bias on the \pinocchio\ catalogs by applying the same mass and modes binning we used in our calibration and measured the correlation matrix between different simulations. In line with the results presented in~\citet{Euclid:2022txd} for the two-point correlation function, we explicitly verified in our analysis that the off-diagonal terms are sub-dominant and of the order of $10$ percent, validating our calibration likelihood.

\subsubsection{\label{sec:hf}Impact of the halo finder on the peak-background split performance}

\begin{figure}
    \vspace{-5pt}
    \centering
    \includegraphics[width=0.99\columnwidth]{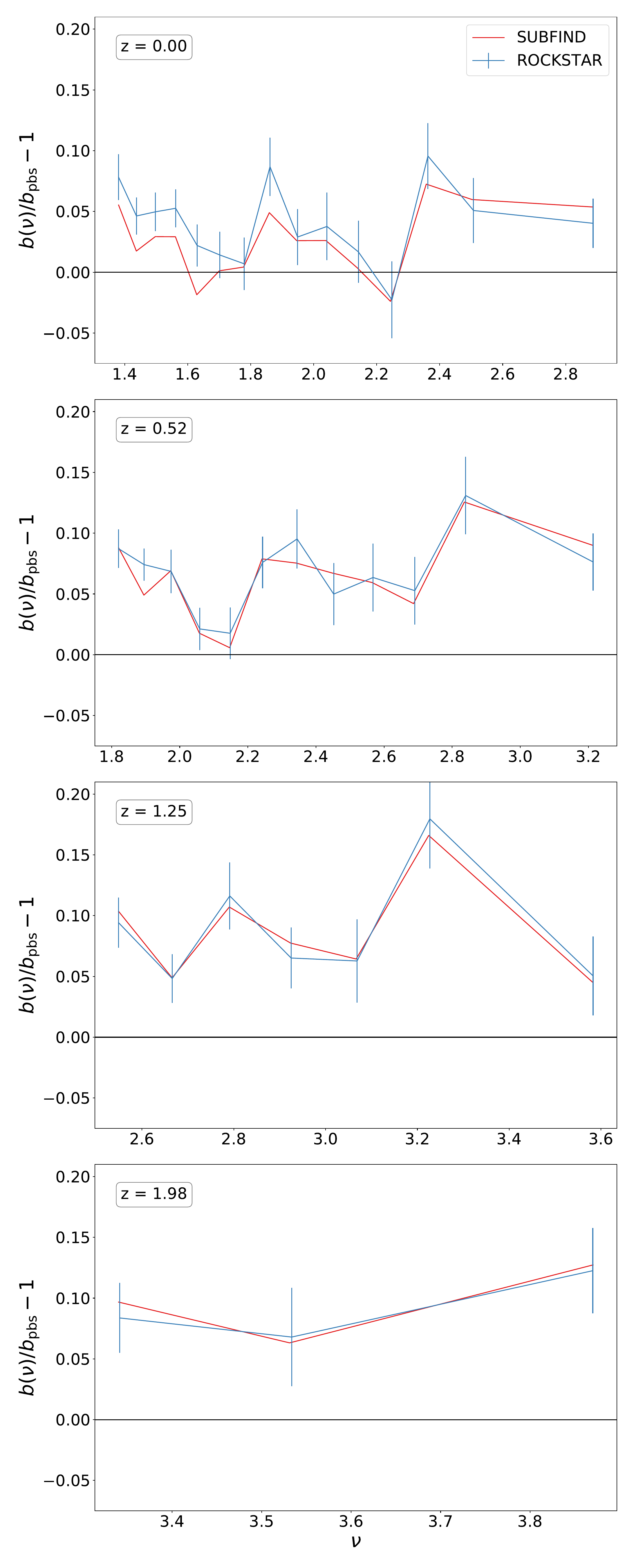}
    \caption{The relative difference between the bias measured in halo catalogs and the bias predicted by the PBS model of Eq. (\protect\ref{eq:biasps}), at different redshifts in the range $z \in [0,2]$. Results refer to simulations carried out for the $C0$ cosmology. In each panel,  blue and red lines refer to the results obtained for the halo catalogs base on the application of  \rockstar\ and \subfind, respectively. For the \rockstar\ catalogs, we also show the standard error of the mean using other $19$ realizations of the same cosmology. }
    \label{fig:pbs-hf}
\end{figure}

In Fig.~\ref{fig:pbs-hf}, the impact of the halo finder on the PBS is examined through the bias ratio measured in halo catalogs generated from the same simulation using either \rockstar\ or \subfind, alongside the corresponding PBS prescription. For the \rockstar\ catalogs, the standard error of the mean is further illustrated using an additional $19$ realizations, with the assumed cosmology being $C0$. The analysis spans different redshifts within the $z \in [0,2]$ range. It is observed that the PBS tends to underestimate the simulation-derived bias by approximately $10\%$ at higher redshifts, though it shows improved accuracy at $z=0$. Given the minimal impact of the choice of halo finder on PBS's overall accuracy, subsequent results will focus exclusively on the \rockstar\ halo catalogs. Despite PBS's tendency to underpredict the HB compared to simulations, it is noteworthy that the deviation remains consistently within 5--15 percent across all explored values of $\nu$ and redshifts. Our future efforts will aim to refine the PBS model by addressing these discrepancies, with the objective of achieving a simulation-calibrated HB model that is precise to within a few percent.

\subsubsection{\label{sec:model}Modelling the halo bias}

\begin{figure*}
    \centering
    \includegraphics[width=\textwidth]{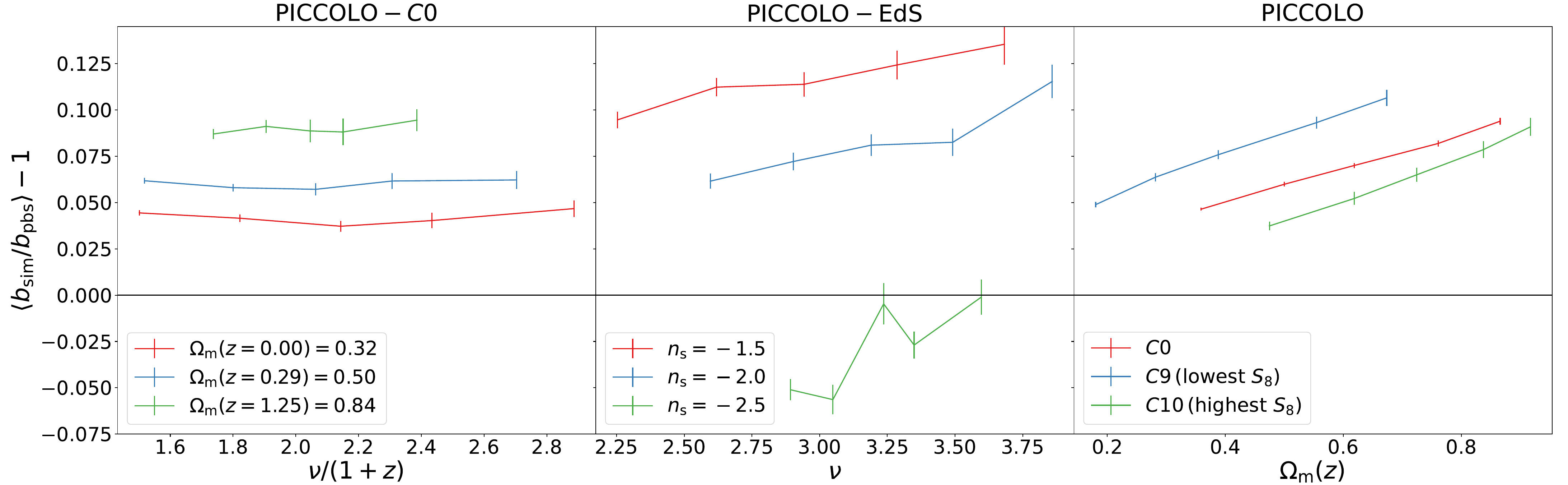}
    \caption{The mean of the ratio of the bias measured in simulations with respect to the PBS prescription. 
    \emph{Left:} the ratio of the bias as a function of $\nu/(1+z)$ for different $z$, labeled by $\Omega_{\rm m}(z)$, for all the $C0$ runs. 
    \emph{Center:} the mean ratio as a function of $\nu$ for the three EdS cosmologies of pure power-law shapes of the linear power spectrum at $z=0$. We report the value of the three spectral indexes in the inset.  
    \emph{Right:} the mean ratio as a function of the background evolution $\Omega_{\rm m}(z)$ for the $C0$, $C9$, and $C10$ cosmologies with varying $S_8$.}
    \label{fig:model}
\end{figure*}

In Fig.~\ref{fig:model}, we present the mean of the ratio of the measured bias with respect to the PBS prescription for different simulations. In the left panel, we use all the $C0$ runs and show the ratio of the bias as a function of $\nu/(1+z)$ for redshifts $0$, $0.29$, and $1.25$. The factor $(1+z)$ is only used to scale the results from different redshifts of the $C0$ model to the same range. In the middle panel, we show the mean ratio as a function of $\nu$ for the three EdS cosmologies. Lastly, in the right panel, we present the mean ratio as a function of the background evolution $\Omega_{\rm m}(z)$ for the $C0$, $C9$, and $C10$ cosmologies.

The left panel of Fig.~\ref{fig:model} illustrates that the performance of PBS is influenced by the cosmological background evolution, encapsulated by $\Omega_{\rm m}(z)$, yet appears unaffected by variations in $\nu$. This contrasts with the observations in the central panel, where PBS performance varies with both $\nu$ and changes in the power spectrum's shape. This sensitivity to $\nu$ is attributed to the limitations of the HMF calibration by~\citet{Euclid:2022dbc} when applied to an EdS cosmology that is far from its calibration regime, introducing artificial dependency.

However, it is important to note that, although these extrapolations to EdS scenarios are beyond the initial calibration range of the HMF model, the model's accuracy is not disproportionately affected across different values of $n_{\rm s}$. Indeed, in~\citet{Euclid:2022dbc}, it has been demonstrated that the HMF model retains a consistent level of precision across various EdS cosmologies characterized by scale-free linear power spectra. Therefore, the dependence on the shape of the power spectrum is more likely related to the varying degree of accuracy of the PBS bias model as the shape of the power spectrum changes.

We interpret the dependence of the PBS performance as a function of the shape of the power spectrum as follows. The extrapolation of the results on the central panel of Fig.~\ref{fig:model} indicates that the PBS performance on EdS cosmologies with a steeper power spectrum ($n_{\rm s}<-2.5$) degrades with decreasing $n_{\rm s}$. Within the PBS framework, the mass variance $\sigma(R_{\rm L})$ smoothed on the scale of the Lagrangian patch $R_{\rm L}$ is assumed to be dominated by the contribution of scales $R_{\rm LSS} \gg R_{\rm L}$. For a power-law power spectrum, it is
\begin{equation}
    \frac{\de \ln{\sigma}}{\de\ln{R}} = -\frac{(n_{\rm}+3)}{2}\,.
\end{equation}
Therefore, the ratio $\sigma(R_{\rm L})/\sigma(R_{\rm LSS})$ tends to unity as $n_{\rm s}$ tends to $-3$. On the one hand, this explains why $n_{\rm s}=-2.5$ presents better performance than the other cases as one of the PBS assumptions is better suited. On the other hand, for $n_{\rm s} \equiv -3$, perturbations on all scales reach the collapse at the same time, and it is no longer possible to distinguish between peaks and a large-scale modulation of a background perturbation, thus breaking PBS's fundamental assumption.

The right panel of Fig.~\ref{fig:model} shows that the residuals with respect to the PBS prediction increase linearly with the value of the density parameter $\Omega_{\rm m}(z)$. While the slope of this linear dependence is similar for the three cosmologies, the normalization is a decreasing function of the clustering amplitude $S_8$. In fact,  $C9$ is the simulation with the lowest $S_8=\left(\sqrt{\Omega_{\rm m,0}/0.3}\,\,\sigma_8\right)= 0.438$ while $C10$ has the highest clustering amplitude $S_8=1.07$. The $C1$ to $C8$ simulations are not shown in this panel for better readability, but they cluster around $C0$ as they have similar $S_8$ values.

The better performance of the PBS in cosmologies with more clustering suggests that the difference between this model and the simulation results is related to the connection between Lagrangian patches in the initial density field and the collapsed structures identified by the halo-finder. Collapsed structures stand out more clearly from the non-linear density field in more evolved and clustered cosmologies. For less clustered models, large halos are still forming and overlapping due to ongoing mergers. This makes it more challenging to identify and link them to their corresponding Lagrangian patches clearly. Not surprisingly, in the EdS cosmologies, the PBS best performance is for $n_{\rm s}=-2.5$, where the evolution of the power spectrum is the steepest. 

The above line of reasoning suggests then that an SO algorithm could not be accurate in providing a one-to-one mapping between Lagrangian patches, destined to form virialized halos according to spherical collapse, and for which the PBS method predicts the bias, and halos identified in the non-linearly evolved density field. In this vein, since both \rockstar\ and \subfind\ are based on an SO algorithm, it is not surprising that they predict similar deviations from PBS (see Fig.~\ref{fig:pbs-hf}).

On the other hand, we expect that collapsed structures have had more time to relax in cosmological models characterized by a higher value of $S_8$. As a consequence, they are more likely to be spherical. Again, this is in line with the better performance of the PBS on evolved cosmologies, as shown in the right panel of Fig.~\ref{fig:model}. Although suggestive, this interpretation of the deviations of PBS predictions would require a dedicated analysis to track their origin in detail, which goes beyond the scope of the analysis presented here.

From the results shown in Fig.~\ref{fig:model}, it emerges that deviations from the PBS should depend on cosmic evolution, parameterized by $\Omega_{\rm m}(z)$, on the slope of the linear power spectrum, and on the clustering amplitude $S_8$.\footnote{There is no fundamental reason for choosing $S_8$ over other parameterizations such as $\sigma_8 \, \left(\Omega_{\rm m, 0}/0.3\right)^\alpha$ with $\alpha$ free to vary. However, we found empirically that $S_8$ works well within our analysis, and thus, we opted to adopt this widely used variable in the literature.} To capture such dependencies, we adopt the following description of the  correction to the PBS prediction for the linear HB
\begin{align}
    \frac{b}{b_{\rm PBS}}&\coloneqq f\left(\Omega_{\rm m}(z), \frac{\de \ln \sigma}{\de \ln R}, S_8\right) \nonumber\\
    &= A_0 \, f_0 \left(\Omega_{\rm m}(z)\right) \, f_1 \left(\frac{\de \ln \sigma}{\de \ln R} \right) \, f_2(S_8)\,.\label{eq:biasmodel}
\end{align}
In the above expression, we assume the following functional forms for the three above dependencies
\begin{align}
    f_0(x) &= 1 + a_1\,x\,,\label{eq:f0}\\
    f_1(x) &= 1 + b_1\,x + b_2\,x^2\,,\\
    f_2(x) &= 1 + c_1\,x\,.\label{eq:f2} 
\end{align}
The parameters $A_0$, $a_1$, $b_1$, $b_2$, and $c_1$ are calibrated in the next section through a detailed comparison with simulations. A balance between simplicity and empirical accuracy drives the parameterization chosen for these contributions. Specifically, we opted for a linear relationship for $\Omega_{\rm m}(z)$ and $S_8$ while modeling the shape of the power spectrum using a quadratic function. In order to assess the potential parameter redundancy in using an extra parameter for the shape of the power spectrum, we performed Watanabe-Akaike Information Criterion (WAIC)~\citep{watanabe2010} and Pareto Smoothed Importance Sampling Leave-One-Out Cross-Validation (PSIS-LOO)~\citep{vehtari2017} analyses comparing the model with $b_2$ free to vary with respect to $b_2$ fixed to $0$. Both analyses confirmed that the fewer degrees of freedom adopted by the surrogate model do not compensate for the decrease in the model's predictability.

Lastly, we report that we do not observe any significant correlation between the model prediction residual and other cosmological parameters assumed in the simulations. Thus, we conclude that the three $f_i$ components in Eq.~\eqref{eq:biasmodel} used in our analysis are sufficient to achieve our goal.

\subsection{\label{sec:cal}Calibration of the halo bias correction}

\begin{figure*}
    \centering
    \includegraphics[width=\textwidth]{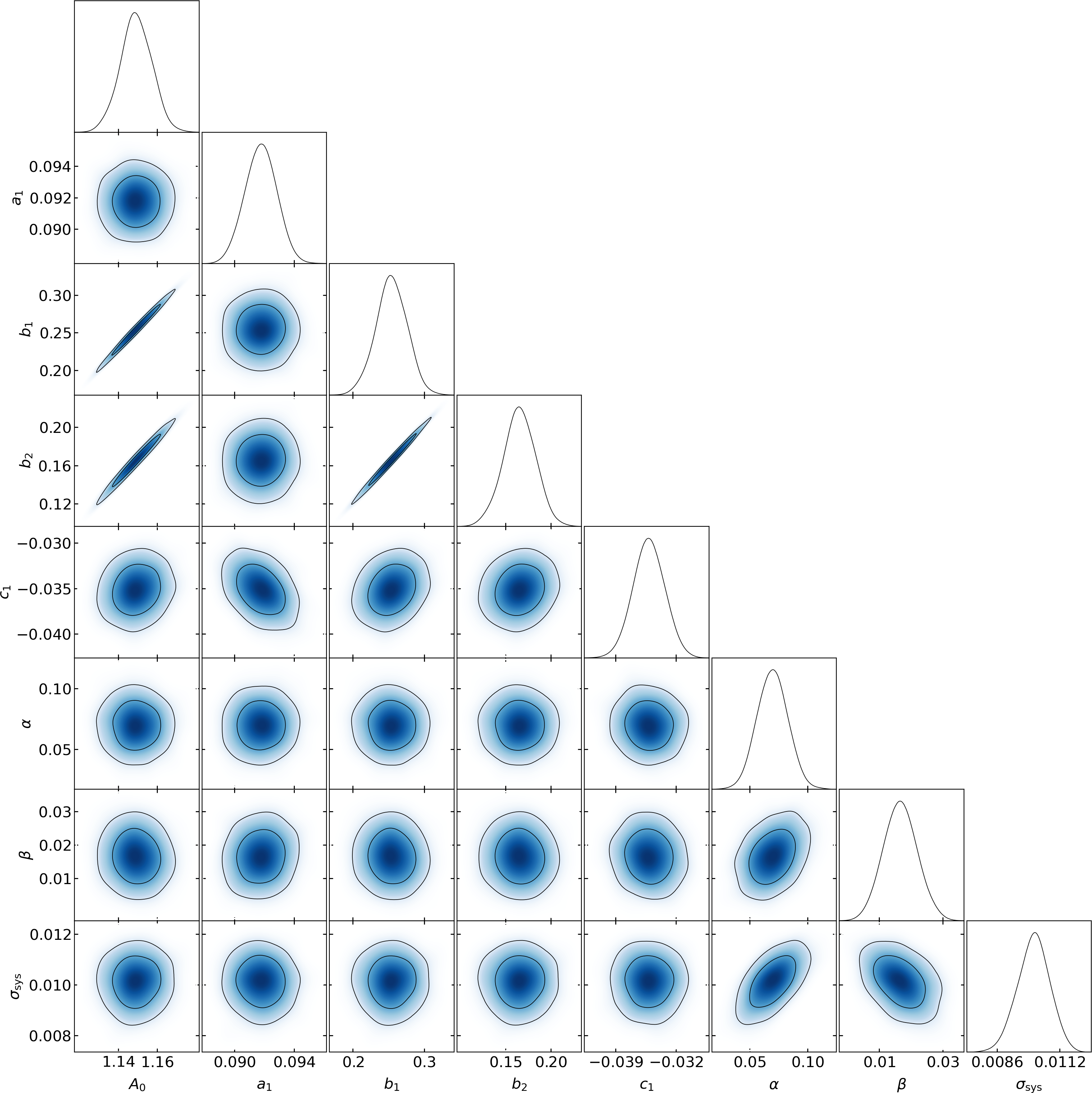}
    \caption{Marginalized 68 and 95 confidence level contours on the model parameters presented in Eqs.~\eqref{eq:biasmodel} to~\eqref{eq:f2}. We calibrate our model using the sub-set of \piccolo\ simulations $C0$--$C10$. See Table~\ref{tab:cal} for the best fit and confidence levels.}
    \label{fig:cal}
\end{figure*}

In Fig.~\ref{fig:cal}, we present the marginalized two-dimensional and uni-dimensional constraints on the model parameters presented in Eqs.~\eqref{eq:biasmodel} to~\eqref{eq:f2}. The best fit and $95$ percent limits are reported in Table~\ref{tab:cal}. We calibrate our model using the sub-set of 60 \piccolo\ simulations covering the cosmological parameters from $C0$ to $C10$.

\begin{table*}[]
    \centering
    \caption{The best fit and $95$ percent limits on the model parameters presented in Eqs.~\eqref{eq:biasmodel} to~\eqref{eq:f2}.}
    \label{tab:cal}
    \resizebox{\textwidth}{!}{
    \renewcommand{\rmdefault}{\small}
    \renewcommand{\arraystretch}{1.5}
    \begin{tabular}{c|c|c|c|c|c|c|c|c|}
    \cline{2-9}
    \multicolumn{1}{c|}{\hspace{0.5em}} & $A_0$ & $a_1$ & $b_1$ & $b_2$ & $c_1$ & $\alpha$ & $\beta$ & $\sigma_{\rm sys}$ \\\hline
     \multicolumn{1}{|c|}{$95\%$ limits}  & $1.149^{+0.016}_{-0.017}$ & $0.0918^{+0.0021}_{-0.0021}$ & $0.254^{+0.042}_{-0.045}$ & $0.165^{+0.034}_{-0.037}$ & $-0.0351^{+0.0036}_{-0.0036}$ & $0.070^{+0.027}_{-0.026}$ & $0.017^{+0.011}_{-0.010}$ & $0.0101^{+0.0013}_{-0.0013}$ \\\hline
     \multicolumn{1}{|c|}{Best fit} & $1.150$ & $0.0929$ & $0.256$ & $0.173$ & $-0.0372$ & $0.071$ & $0.012$ & $0.0107$\\\hline\hline
    \end{tabular}
    }
    \tablefoot{We calibrate our model using the sub-set of \piccolo\ simulations $C0$--$C10$.}
\end{table*}

\begin{figure*}
    \centering
    \includegraphics[width=\textwidth]{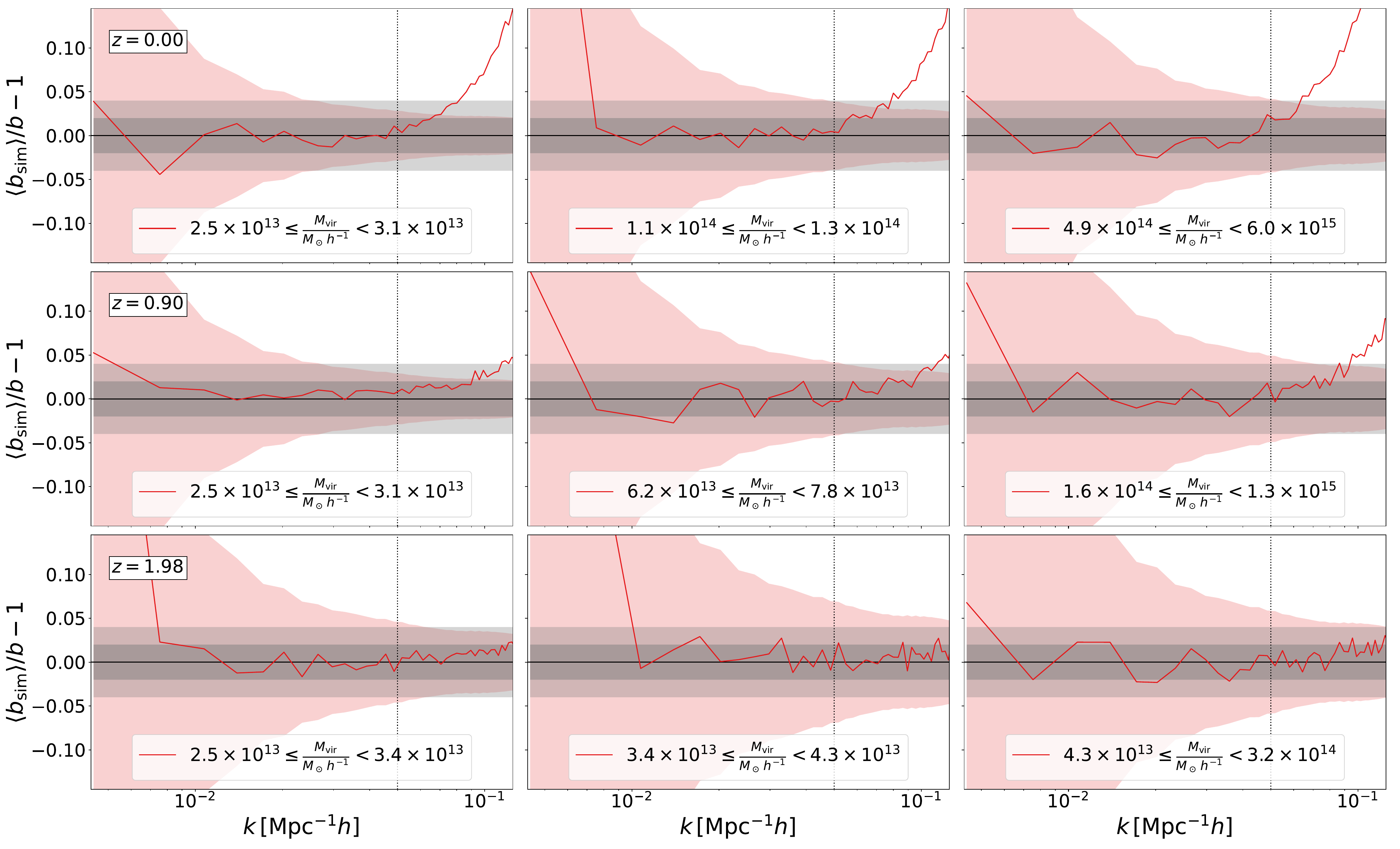}
    \caption{The ratio between the mean of the observations on the 20 $C0$ simulations with respect to our model predictions. Different rows correspond to different redshifts, while each panel corresponds to a different mass bin. The shaded region in red corresponds to the error on the mean, assuming that each measurement follows Eq.~\eqref{eq:cross_var_2}. The shaded regions in grey correspond to two and four percent regions.}
    \label{fig:C0}
\end{figure*}

In Fig.~\ref{fig:C0}, we present the ratio between the mean of the observations on the 20 $C0$ simulations with respect to our model predictions. Different rows correspond to different redshifts, while each panel corresponds to a different mass bin. The presented mass bins were selected as before to span from the least massive to the most massive occupied bin in that redshift. The shaded region in red corresponds to the error on the mean, assuming that each measurement follows Eq.~\eqref{eq:cross_var_2}. The shaded regions in grey correspond to two and four percent regions. As can be seen, our model's prediction presents a performance below two percent for different masses and redshift regimes when not primarily dominated by the sample variance, as it happens, for instance, for $k\lesssim 10^{-2}\,\mpcinv$. This accuracy holds over the $k$ range up to which the onset of non-linearity occurs and the approximation of scale-independent bias breaks down, i.e., $k\gtrsim 0.05 \,\mpcinv$ (marked by a vertical line). 

We note that at large $k$ values, non-linearity effects cause the bias measured from the simulations to take a scale dependence in exceeding the model prediction. As expected, this effect is smaller at higher redshift, consistent with the effect of non-linearity shifting to larger $k$ values.

\begin{figure*}
    \centering
    \includegraphics[width=\textwidth]{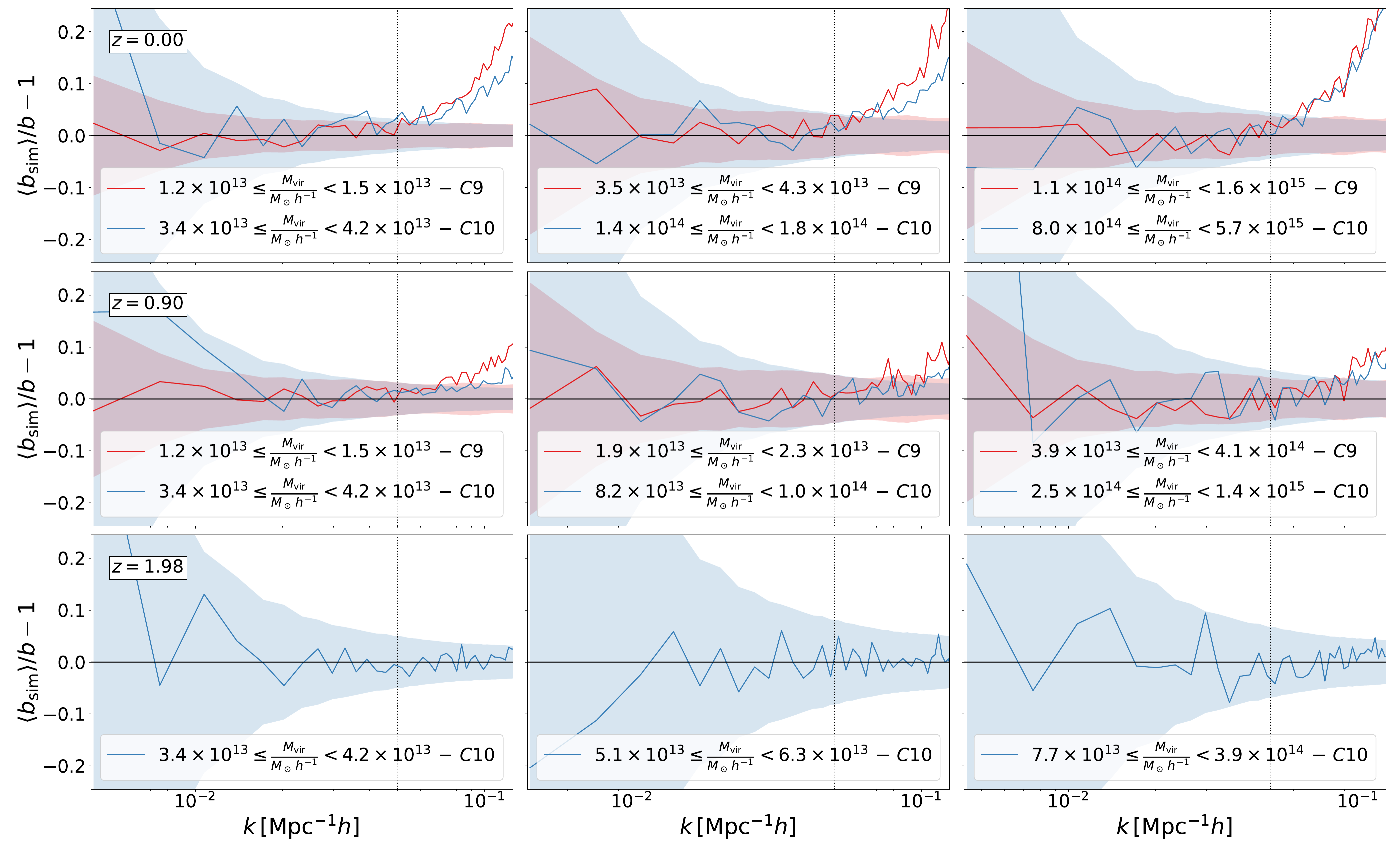}
    \caption{Similar to Fig.~\ref{fig:C0} but for the $C9$ and $C10$ cosmological parameters. Among the \piccolo\ set, $C9$ and $C10$ correspond to the cosmologies with the lowest and the highest $S_8$, respectively.}
    \label{fig:C9C10}
\end{figure*}

Similarly to Fig.~\ref{fig:C0}, in Fig.~\ref{fig:C9C10}, we present the ratio between the mean of the observations on the $C9$ and $C10$ simulations with respect to our model predictions. $C9$ and $C10$ are the simulations with lowest and highest $S_8$, respectively. Even for such extreme scenarios, our model performs well, thus confirming that our linear bias model, with the previously described calibration, can reproduce results from simulations with an accuracy of a few percent for $\Lambda$CDM cosmologies. 

\subsection{\label{sec:neutrinos}Cosmologies with massive neutrinos}

\begin{figure*}
    \centering
    \includegraphics[width=\textwidth]{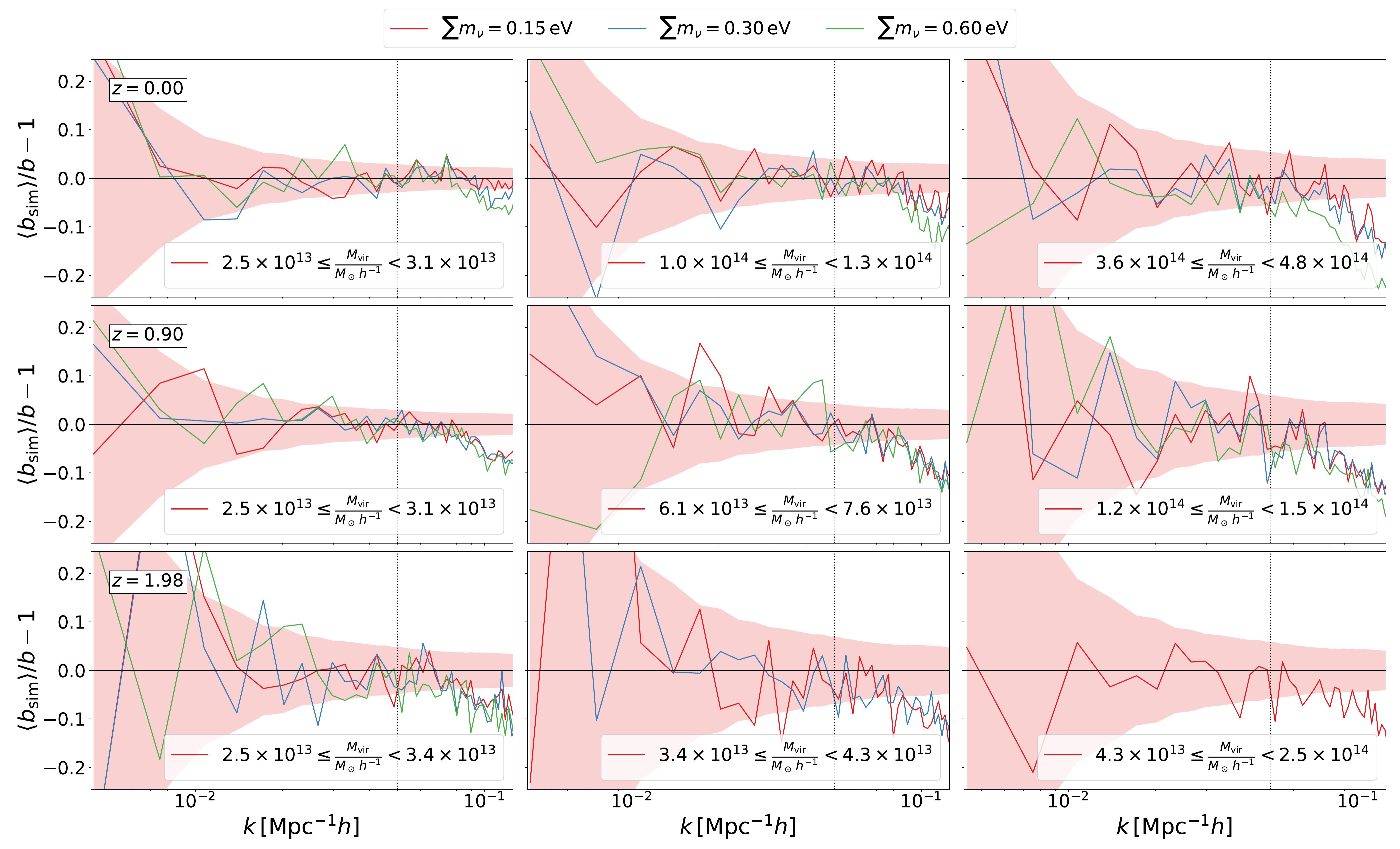}
    \caption{Similar to Fig.~\ref{fig:C0} but for simulations with massive neutrinos. For better plot readability, we only show the uncertainties (red shaded regions) for the simulation with total neutrino masses equal to $0.15\,{\rm eV}$.}
    \label{fig:C0-nus}
\end{figure*}

We present our model's performance when considering simulations with massive neutrinos in Fig.~\ref{fig:C0-nus}. This allows us to assess the performance of our HB calibration for this minimal extension of $\Lambda$CDM. In this case, the simulation's bias has been computed by comparing it to the linear power spectrum of the corresponding cosmological model, which includes only the contributions from cold dark matter and baryons. For consistency, the same choice of considering only CDM and baryon contribution is also made for the computation of the HMF entering in our model for the HB~\citep[see, for instance,][]{Castorina:2014,Costanzi:2013bha}. From Fig.~\ref{fig:C0-nus}, it is evident that our model also precisely describes the bias in $\Lambda (\nu)$CDM models, despite the fact that such models have not been used during the HB calibration.

We note that, unlike for the pure $\Lambda$CDM models, in this case, the measured bias underpredicts the model bias at large $k$. We also note that the dependence of this effect on redshift, if any, goes in the direction of being larger at higher $z$. Also, there is some hint for it to be slightly smaller for smaller values of $m_\nu$ and, therefore, of $\Omega_\nu$. These effects align with the expectation that such deviations are not dominated by non-linear evolution but rather by the effect of neutrino-free streaming~\citep{Castorina:2014}.

\subsection{\label{sec:comparing}Comparison with previous models}
\begin{figure*}
    \centering
    \includegraphics[width=\textwidth]{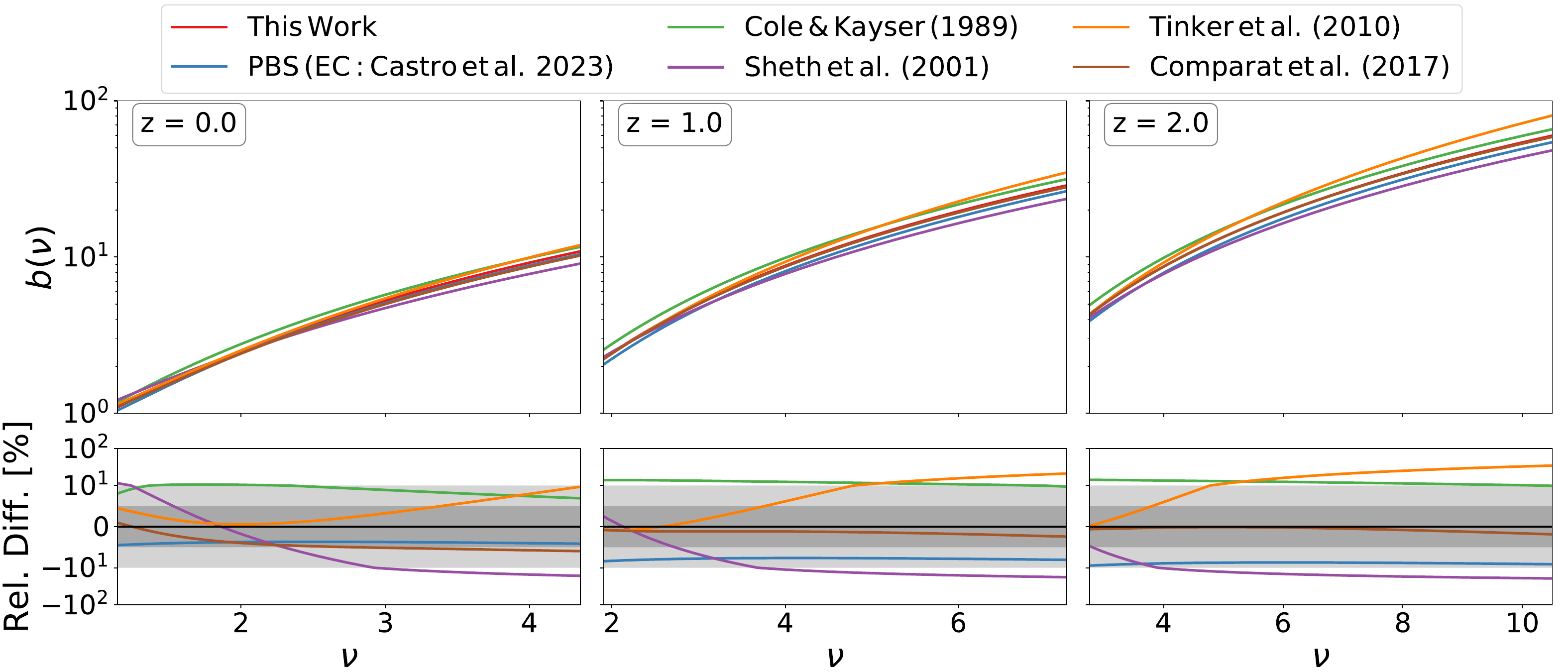}
    \caption{Comparison between the HB predicted by our model with predictions from other models presented in the literature:~\citet{Cole:1989vx},~\citet{Sheth:1999su},~\citet{Tinker:2010my}, and~\citet{Comparat:2017ejl}. We present both our benchmark model as well as the PBS predictions based on the HMF model of~\citet{Euclid:2022dbc} used as a baseline of our model.  Different columns correspond to different redshifts. The relative difference with respect to our benchmark model is presented in the panels in the second row. We adopt a composite scale for the residual plot to show the dynamic range of differences between the models: the scale is linear for values between $[-10, 10]\,\%$ and symmetric log outside. For reference, we show the zero line in black. The predictions of the models from the literature have been computed using the COLOSSUS toolkit~\citep{Diemer:2017bwl}. }
    \label{fig:comp}
\end{figure*}

In Fig.~\ref{fig:comp}, we compare our model prediction with other models in the literature:~\citet{Cole:1989vx},~\citet{Sheth:1999su},~\citet{Tinker:2010my}, and~\citet{Comparat:2017ejl}. We present both our benchmark model as well as the PBS predictions based on the HMF model of~\citet{Euclid:2022dbc}, which we use as a baseline of our model.  Different columns correspond to different redshifts. The relative difference to our benchmark model is presented in the panels in the second row. The predictions of the external models have been computed using the COLOSSUS toolkit~\citep{Diemer:2017bwl}.\footnote{\url{https://bdiemer.bitbucket.io/colossus/}} 

To ensure a fair comparison, we adopt the Planck-like $C0$ cosmology, where the compared models have their peak performance among the \piccolo\ cosmologies. All compared models have degraded performance as we move from this benchmark cosmology while we have shown the robustness of our model in Figs.~\ref{fig:C0}, ~\ref{fig:C9C10}, and~\ref{fig:C0-nus}. That is due to the models assuming either the universality of the bias relation or a redshift-only dependence, while our method explicitly models the cosmology dependence of this relation. As such, comparisons between these models should be interpreted cautiously, as the underlying cosmology influences the exact figures. 

As for the comparison with the PBS prediction based on the HMF calibration by \citet{Euclid:2022dbc}, the results shown here confirm those shown in Fig.\ref{fig:model}: the PBS-based predictions underestimate our simulations-based calibration by about 5--10\%, almost independently of $\nu$.
The models of~\citet{Cole:1989vx} and~\citet{Sheth:1999su} over- and under-estimate the bias by $\sim 10 \%$, respectively. Their relatively poor performance is not surprising. The prediction by \citet{Cole:1989vx} corresponds to the PBS prediction when using the HMF from ~\citet{Press:1973iz}. As the~\citet{Press:1973iz} HMF only qualitatively explains the abundance of halos, it is expected that the bias from the PBS will not perform much better. On the other hand, the~\citet{Sheth:1999su} model was calibrated on simulations. However, such simulations had a resolution that allowed those authors to cover a dynamic range significantly narrower than that accessible to our simulations. The prediction by \citet{Cole:1989vx} differs from ours by an amount almost independent of $\nu$ and redshift. On the other hand, the HB by \cite{Sheth:1999su} differs from ours in a $\nu$- and $z$-dependent way. 

Notably, the model of~\citet{Tinker:2010my} is only superior to the abovementioned models for low peak-height. The differences with respect to our model grow with redshift and peak-height. This could be due to the heterogeneity of the simulations used to calibrate the model of~\citet{Tinker:2010my} and the possible limitation of the model itself to capture the cosmological dependence of the HB. In this paper, we calibrate to a set of simulations that have been run with the same code and setup. On the other hand,~\cite{Tinker:2010my} based their analysis on a collection of simulations carried out with different codes and configurations in terms of resolution and box sizes. Also, their model assumes a redshift dependency for the evolution, while from Fig.~\ref{fig:cal} we note that a parametrization of this evolution through $\Omega_{\rm m}(z)$ is more universal.

Lastly, the model by~\citet{Comparat:2017ejl} shows a good agreement with our model. The most significant differences are at low-redshift where the model of~\citet{Comparat:2017ejl} predicts a bias that is smaller than ours by about $5$ percent. This difference reduces to a sub-percent at high redshift. The agreement is unsurprising as the model of~\citet{Comparat:2017ejl} was also calibrated on \rockstar\ catalogs.

\subsection{\label{sec:cc}Impact on cluster cosmology analysis}

In this section, we forecast the impact of the HB calibration on cosmological analyses of cluster counts and cluster clustering from \Euclid. We present the results for the bias model of \citet{Tinker:2010my} and the model calibrated in this study. The rationale for assuming~\citet{Tinker:2010my} is to use a widely used model in cluster cosmology representative of the difference in the bias models presented in Fig.~\ref{fig:comp}. Nonetheless, we do not expect the results to change significantly if we assumed the model of~\citet{Comparat:2017ejl} that presents a better concordance with our model at high redshift but worse at low redshift. Therefore, the overall impact on cosmological constraints will compensate partially as the clustering cosmological signal for \Euclid\ peaks at lower redshifts. 

As described in Sect.~\ref{sec:forecast}, we assess the effect of the HB calibration in a more realistic scenario, performing a likelihood analysis of both the individual analysis of number counts and cluster clustering statistics and the combination of these probes. In all scenarios, the observable-mass relation (Eq.~\ref{eq:lambda}) is calibrated by combining the probes with weak lensing (WL) mass estimates, assuming a constant error of 1\%. The mass calibration is the primary source of systematic uncertainty in cluster cosmology studies, and a 1\% calibration represents the goal for stage IV surveys. Therefore, the chosen setting offers a forecast of the utmost cosmological bias resulting from inaccurate modeling of the HB. Lastly, we assume three independent Gaussian likelihoods \citep{Fumagalli:2023yym} for number counts, clustering, and WL masses.

\begin{figure*}[h]
    \centering
    \includegraphics[width=\textwidth]{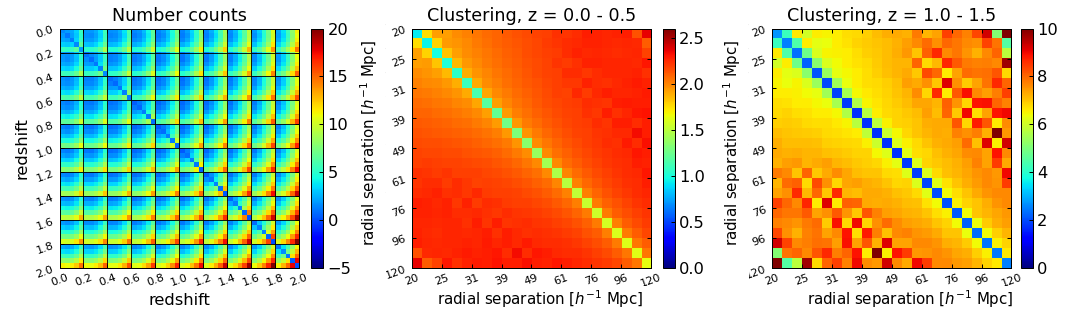}
    \caption{Percentage residuals of cluster counts (\textit{left} panel) and cluster clustering covariance matrices (\textit{central} and \textit{right} panels), computed with the bias from \citet{Tinker:2010my} in comparison to the one calculated using the bias calibrated in this study (Eq.~\ref{eq:biasmodel}). We show the full covariance matrix for number counts in mass and redshift bins. For the two-point correlation function of galaxy clusters, we show two blocks of the full covariance (low and high redshift bins) as a function of the radial separation.}
    \label{fig:cosmo_cov}
\end{figure*}

In the left panel of Fig.~\ref{fig:cosmo_cov}, we start by presenting the percentage residuals of the number counts covariance. We show the full covariance matrix for the number counts analysis, with the mass dependence within each redshift bin. Notably, the impact of the HB model is minimal at low redshift but becomes significant, reaching up to 20\%, at higher redshifts, especially in the high-mass bins. However, the impact of a different bias calibration is mitigated by the shot-noise contribution when the latter becomes dominant along the diagonal, as the HB only plays a role in the computation of sample variance. To quantify the impact of such a discrepancy on parameter posteriors, we perform the likelihood analysis for a number counts experiment, as described in Sect.~\ref{sec:forecast}. From the comparison of the two posteriors, we obtain $\Delta {\rm FOM} = -0.67$ and a perfect agreement between the positions of the two contours, meaning that the impact of the HB calibration is below other systematics.

As for the analysis of cluster counts, in the central and right panels of Fig.~\ref{fig:cosmo_cov}, we present the percent residuals for the clustering covariance as a function of the radial separation in both a low-redshift (central panel) and high redshift (right panel) bin. Similar to our findings for the number counts, the most significant impact is observed on the off-diagonal elements, particularly at high redshift. However, in contrast to number counts covariance, the inclusion of shot-noise in cluster clustering contributes to the off-diagonal elements, helping to mitigate the effect of different HB calibrations across all scales. This results in a difference that never exceeds 10\%. Importantly, in the case of cluster clustering, the HB also affects the expected signal -- either the two-point correlation function or power spectrum -- leading to a difference independent of the radial separation but increasing with redshift, reaching a 10\% level in the high-redshift interval. 

The cosmological forecasts from the clustering experiment show that the minimal variation in covariance terms produces a negligible difference in the posterior amplitude, equal to $\Delta {\rm FOM} = 1.03$. However, comparing the two posteriors reveals an agreement at only $0.68\,\sigma$. This implies that the differences in the two-point correlation function translate into a sizeable shift in the cosmological constraints. Notably, the difference in the posteriors induced by the different calibrations of the HB alone surpasses the $0.25\,\sigma$ threshold commonly employed in other studies~\citep{Euclid:2023wdq} to flag systematic errors that, if exceeded, could accumulate and lead to a collectively significant difference.

\begin{figure}[h]
    \centering
    \includegraphics[width=0.49\textwidth]{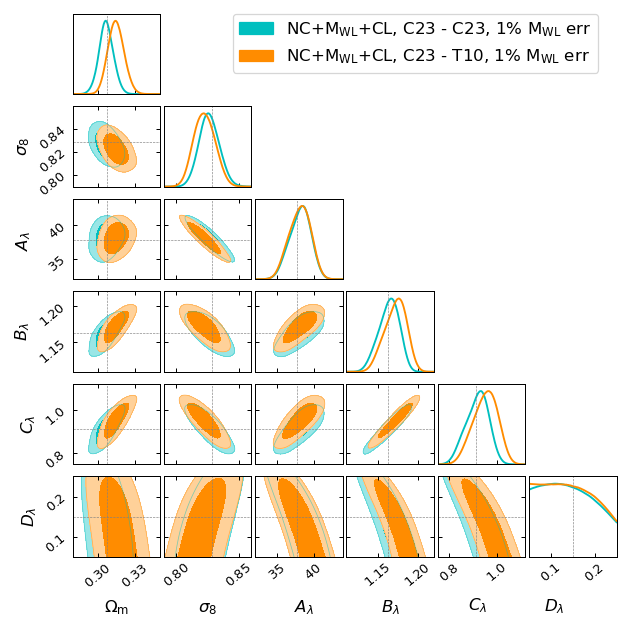}
    \caption{Parameter posteriors at 68 and 95 percent of confidence level obtained by analyzing number counts, weak lensing masses, and cluster clustering computed with the HB calibrated in this work (cyan contours) and the bias from \citet{Tinker:2010my} (orange contours). The error associated with weak lensing mass is set at 1\% of the mass.}
    \label{fig:cosmo_post}
\end{figure}

The combined analysis (cluster counts + cluster clustering) posteriors are shown in Fig.~\ref{fig:cosmo_post}. We notice that assuming the HB calibration by \citet{Tinker:2010my} still causes a shift in the posteriors with respect to the HB calibration presented in this paper. This aligns with the forecast results for the cluster clustering analysis presented above. Although the difference is reduced to $0.39\,\sigma$, the combination with number counts and weak lensing masses cannot compensate for the impact of the HB calibration that affects mostly the cluster clustering. 


\section{\label{sec:conclusions}Conclusions}

This paper presents a calibrated semi-analytical model for the HB in view of the joint cosmological exploitation of number counts and clustering of galaxy clusters from the Euclid survey. Our approach begins with the PBS model, based on the HMF of \citet{Euclid:2022dbc}, and extends this approach by introducing a novel parametric correction. This correction is designed to align the PBS prediction with the results from an extended and homogenous set of $N$-body simulations that we carried out for both vanilla $\Lambda$CDM models, varying cosmological parameters, and for $\Lambda(\nu)$CDM models by varying the sum of the neutrino masses. The simulations employed fixed and paired initial conditions~\citep[see,][]{Angulo:2016hjd}, providing a robust reduced variance framework for our analysis. We measured the HB by examining the ratio of the matter-halo cross-spectrum. Additionally, we modeled the covariance of these measurements using 200 mock catalogs of the Euclid cluster survey, based on the approximate LPT-based \pinocchio\ code. This ensures a thorough understanding of the uncertainties involved in our calibration of the HB. The key findings and implications of our study are summarized in the following points.

\begin{itemize}
    \item The use of fixed and paired initial conditions for the simulations analysed in our study proved highly advantageous for estimating the bias of tracers. By parametrizing the covariance of the bias measurements with two parameters -- one controlling the shot-noise contribution and the other for suppression due to fixing, respectively $\alpha$ and $\beta$ in Eq.~\eqref{eq:cross_var_2} -- we observed significant effectiveness in the variance suppression term. This was demonstrated in the constraints on the terms describing the variance in the halo-matter power spectrum, $P_{\rm hm}(k)$, shown in Fig.~\ref{fig:f&pchain}. Furthermore, our analysis of the measurements of $P_{\rm hm}(k)$ between paired simulations, as illustrated in Fig.~\ref{fig:f&pcor}, revealed no significant correlation between them. This finding underscores the efficacy of the fixed and paired simulation approach in providing reliable estimates of the bias factor characterizing the distribution of tracers (i.e., halos), which is free from the influences of inherent correlations that could affect the results.
    
    \item The impact of the choice of the halo finder used in the analysis of the $N$-body simulations on the performance of the PBS is shown in Fig.~\ref{fig:pbs-hf}. While comparing the \rockstar\ and \subfind\ halo finders, it was observed that the PBS generally underestimates the measured bias from simulations, particularly at higher redshifts. However, the impact of the halo finder choice on the PBS prescription's performance is almost negligible, thus reinforcing the robustness of our approach in assessing the PBS performance across different redshift ranges.

    \item Our modeling of the HB, as illustrated in Fig.~\ref{fig:model}, reveals significant insights into the cosmological dependency of the performance of the PBS. The background cosmological evolution influences the PBS performance more than the peak height parameter $\nu$. This was particularly evident in different cosmologies, where PBS's effectiveness varied with the shape of the power spectrum and the degree of clustering evolution, as described by the $S_8$ parameter. Notably, in more clustered cosmologies, PBS improved its performance. This suggests a possible link between the ease of identifying collapsed structures in cosmologies with more evolved clustering and their corresponding Lagrangian patches. This result led us to develop a refined model for the PBS correction, expressed in Eq.~\eqref{eq:biasmodel}, which incorporates terms depending on $\Omega_{\rm m} (z)$, the local slope of the power spectrum, and $S_8$. 

    \item The calibration of our model parameters, with best-fit presented in Fig.~\ref{fig:cal} and Table~\ref{tab:cal}, demonstrated its robust performance across a range of cosmological conditions. Figures~\ref{fig:C0} and \ref{fig:C9C10} illustrate our model prediction performance on the reference $C0$ simulations and on the $C9$ and $C10$ simulations. The accuracy of our model is particularly noteworthy, always remaining below a two percent deviation for different masses and redshift regimes, with the possible exception of unrealistic cases largely influenced by sample variance. Quite remarkably, this level of precision is maintained even in extreme scenarios represented by the $C9$ and $C10$ simulations, which have the lowest and highest $S_8$ values, respectively. 

    \item The robustness of our HB calibration is further demonstrated in scenarios involving massive neutrinos, as showcased in Fig.~\ref{fig:C0-nus}. Despite not incorporating massive neutrino simulations during the calibration phase, our model accurately predicts the HB in these cosmologies. Neutrinos are treated according to the model presented by~\citet[][see also \citealt{Costanzi:2013bha}]{Castorina:2014} and the measurements of the bias with respect to the matter power spectrum of cold dark matter and baryons, as it was done for the HMF in simulations with massive neutrinos in~\citet{Euclid:2022dbc}. The ability of our model to adapt and perform reliably in such scenarios without the need for recalibration highlights its robustness and versatility.

    \item As for the comparison with HB models already introduced in the literature (see Figure~\ref{fig:comp}), the models by \citet{Cole:1989vx} and \citet{Sheth:1999su} show significant deviations from our results, likely due to their calibration on simulations covering narrower dynamic ranges and variety of cosmological models. The HB model by \citet{Tinker:2010my} shows increasing discrepancies with our model at higher redshifts and peak heights. Such differences could be attributed to its calibration on a heterogeneous set of simulations and an inadequate accounting for the cosmological dependence of the HB. In contrast, the model by \citet{Comparat:2017ejl} aligns more closely with our findings, particularly at higher redshifts. This agreement is expected as their model was also calibrated using \rockstar\ catalogs. 

    \item As for the impact of changing the calibration of the HB on the derived cosmological posteriors, we showed in Fig.~\ref{fig:cosmo_cov} the differences of the covariance matrices for a \Euclid cluster count and cluster clustering analysis using both our calibration and the one provided by \cite{Tinker:2010my}. While the impact on number counts covariance is minimal at low redshifts, it becomes substantial, up to 20\%, at higher redshifts. However, the presence of shot-noise in the analysis helps mitigating this effect. In cluster clustering, we observed that the HB calibration could lead to differences in the two-point correlation function, particularly at high redshifts. This difference can potentially bias cosmological constraints beyond the $0.25\,\sigma$ threshold commonly used to flag significant systematic errors~\citep{Euclid:2022qde}. Moreover, the combined analysis of number counts, cluster clustering, and weak lensing masses demonstrates that even with these additional data, the calibration of HB can not be entirely compensated. This highlights the importance of precise HB calibration in cluster cosmology, especially for a survey like the one being provided by \Euclid that will reach an unprecedented sensitivity and level of statistics.
\end{itemize}

In summary, the analysis presented in this paper has systematically calibrated and tested the HB for a range of cosmological scenarios, demonstrating its critical impact on the precision of cosmological analyses based on galaxy clusters for the \Euclid mission. The resilience of our HB model against variations of cosmological models, including the presence of massive neutrinos and different degrees of clustering amplitude, highlights its robustness and adaptability. Importantly, our model is robust against the halo finder definition, inheriting its dependence through the HMF only. This is a remarkable feature as the correspondence between halos in \textit{N}-body simulations and real clusters in surveys remains a complex issue, with uncertainties in halo identification and characterization potentially influencing the extraction of cosmological parameters. Future research should focus on understanding and quantifying these uncertainties, especially concerning observational challenges such as projection effects and the mass-observable relation. As we move forward, extending this precision to departures from the standard $\Lambda(\nu)$CDM framework will be crucial in fully harnessing the capabilities of next-generation cosmological surveys.

\section{Data availability}
\label{sec:dataavailability}
In~\citet{Castro_CCToolkit_A_Python_2024}, we implement the model presented in this paper, together with the models for the HMF presented in~\citet{Euclid:2022dbc} and for the impact of baryonic feedback on cluster masses presented in~\citet{Euclid:2023jih}. The source code can be accessed in~\faGithub~\href{https://github.com/TiagoBsCastro/CCToolkit}{https://github.com/TiagoBsCastro/CCToolkit}.

\begin{acknowledgements}
It is a pleasure to thank Valerio Marra for constructive comments during the production of this work, Fabio Pitari and Caterina Caravita for support with the CINECA environment, Peter Berhoozi for the support with \rockstar, Oliver Hahn for the support with \monofonic, and Luca Di Mascolo for the support with \pymc. TC is supported by the Agenzia Spaziale Italiana (ASI) under - Euclid-FASE D  Attivita' scientifica per la missione - Accordo attuativo ASI-INAF n. 2018-23-HH.0. SB, TC, PM, and AS are supported by the PRIN 2022 PNRR project "Space-based cosmology with Euclid: the role of High-Performance Computing" (code no. P202259YAF), by the Italian Research Center on High-Performance Computing Big Data and Quantum Computing (ICSC), project funded by European Union - NextGenerationEU - and National Recovery and Resilience Plan (NRRP) - Mission 4 Component 2, within the activities of Spoke 3, Astrophysics and Cosmos Observations, and by the INFN INDARK PD51 grant. TC and AS are also supported by the FARE MIUR grant `ClustersXEuclid' R165SBKTMA. AS is also supported by the ERC `ClustersXCosmo' grant agreement 716762. MC and TC are supported by the PRIN 2022 project EMC2 - Euclid Mission Cluster Cosmology: unlock the full cosmological utility of the Euclid photometric cluster catalog (code no. J53D23001620006). KD acknowledges support by the DFG (EXC-2094 - 390783311) as well as support through the COMPLEX project from the European Research Council (ERC) under the European Union’s Horizon 2020 research and innovation program grant agreement ERC-2019-AdG 882679. We acknowledge the computing centre of CINECA and INAF, under the coordination of the ``Accordo Quadro (MoU) per lo svolgimento di attività congiunta di ricerca Nuove frontiere in Astrofisica: HPC e Data Exploration di nuova generazione'', for the availability of computing resources and support. We acknowledge the use of the HOTCAT computing infrastructure of the Astronomical Observatory of Trieste -- National Institute for Astrophysics (INAF, Italy) \citep[see][]{2020ASPC..527..303B,2020ASPC..527..307T}.  \AckEC
\end{acknowledgements}

%
%

\bibliography{euclid}

%

\begin{appendix}
\end{appendix}

\end{document}